\documentclass[10pt,journal,compsoc]{IEEEtran}

\ifCLASSOPTIONcompsoc
  \usepackage[nocompress]{cite}
\else
  \usepackage{cite}
\fi
\usepackage[pdftex]{graphicx}
\usepackage{amsmath,amssymb,amsfonts}
\usepackage{algorithmic}
\usepackage{algorithm}
\usepackage{array}
\usepackage{url}
\usepackage{color}
\usepackage{hyperref}
\usepackage{dsfont}
\usepackage{bm}
\usepackage{subfigure}
\usepackage{float}
\renewcommand{\vec}[1]{\boldsymbol{#1}}
\DeclareMathOperator*{\argmin}{argmin}

\newtheorem{proposition}{Proposition}[section]
\newtheorem{definition}{Definition}[section]
\newtheorem{proof}{Proof}[section]

\begin{document}

\title{Distributed Redundant Placement for Microservice-based Applications at the Edge}

\author{Hailiang~Zhao,
        Shuiguang~Deng,~\IEEEmembership{Senior~Member,~IEEE,}
        Zijie~Liu,
        Jianwei~Yin,
        and~Schahram~Dustdar,~\IEEEmembership{Fellow,~IEEE}
\IEEEcompsocitemizethanks{\IEEEcompsocthanksitem H. Zhao, S. Deng, Z. Liu, and J. Yin are with 
the College of Computer Science and Technology, Zhejiang University, Hangzhou 310058, China.\protect\\
E-mail: \{hliangzhao, dengsg, liuzijie, zjuyjw\}@zju.edu.cn
\IEEEcompsocthanksitem S. Dustdar is with the Distributed Systems Group, Technische Universität Wien, 1040 Vienna, Austria.\protect\\
E-mail: dustdar@dsg.tuwien.ac.at
\IEEEcompsocthanksitem Shuiguang Deng is the corresponding author.}
}

\IEEEtitleabstractindextext{%
\begin{abstract}
Multi-access Edge Computing (MEC) is booming as a promising paradigm to push the computation and communication 
resources from cloud to the network edge to provide services and to perform computations. With container 
technologies, mobile devices with small memory footprint can run composite microservice-based applications 
without time-consuming backbone. Service placement at the edge is of importance to put MEC from theory into 
practice. However, current state-of-the-art research does not sufficiently take the composite property of 
services into consideration. Besides, although Kubernetes has certain abilities to heal container failures, 
high availability cannot be ensured due to heterogeneity and variability of edge sites. To deal with these 
problems, we propose a distributed redundant placement framework SAA-RP and a GA-based Server Selection (GASS) 
algorithm for microservice-based applications with sequential combinatorial structure. We formulate a stochastic 
optimization problem with the uncertainty of microservice request considered, and then decide for each microservice, 
how it should be deployed and with how many instances as well as on which edge sites to place them. Benchmark 
policies are implemented in two scenarios, where redundancy is allowed and not, respectively. Numerical results 
based on a real-world dataset verify that GASS significantly outperforms all the benchmark policies. 
\end{abstract}

}

\maketitle

\IEEEdisplaynontitleabstractindextext

%
\IEEEpeerreviewmaketitle

\IEEEraisesectionheading{\section{Introduction}\label{sec1}}
\IEEEPARstart{N}{owadays}, mobile applications are becoming more and more computation-intensive, location-aware, 
and delay-sensitive, which puts a great pressure on the traditional Cloud Computing paradigm to guarantee the 
Quality of Service (QoS). To address the challenge, Multi-access Edge Computing (MEC) was proposed to provide 
services and to perform computations at the network edge without time-consuming backbone transmission, so as to 
enable fast responses for mobile devices \cite{MEC_survey1}\cite{MEC_survey2}\cite{review-add1}.

MEC offers not only the development on the network architecture, but also the innovation in service 
patterns. Considering that small-scale data-centers can be deployed near cellular tower sites, there are 
exciting possibilities that microservice-based applications can be delivered to mobile devices without backbone transmission, in virtue of 
setting up a unified service provision platform. Container technologies, represented by Docker \cite{Docker}, 
and its dominant orchestration and maintenance tool, Kubernetes \cite{Kubernetes}, are becoming the mainstream 
solution for packaging, deploying, maintaining, and healing applications. Each microservice decoupled from the 
application can be packaged as a Docker image and each microservice instance is a Docker container. Here we take 
Kubernetes for example. Kubernetes is naturally suitable for building \textit{cloud-native} applications by 
leveraging the benefits of the distributed edge because it can hide the complexity of microservice orchestration 
while managing their availability with lightweight Virtual Machines (VMs), which greatly motivates Application 
Service Providers (ASPs) to participate in service provision within the access and core networks. 

Service deployment from ASPs is the carrier of service provision, which touches on where to place the services 
and how to deploy their instances. In the last two years, there exist works study the placement at the network 
edge from the perspective of Quality of Experience (QoE) of end users or the budget of ASPs \cite{placement-follow-me} 
\cite{placement-2} \cite{placement-3} \cite{placement-MAB} \cite{placement-1}\cite{placement-our-work} \cite{placement-add1}. However, 
those works commonly have two limitations. Firstly, the \textit{to-be-deployed} service only be studied in an atomic 
way. It is often treated as a single abstract function with given input and output data size. Time series or 
composition property of services are not fully taken into consideration. Secondly, high availability of deployed service 
is not carefully studied. Due to the heterogeneity of edge sites, such as different CPU cycle frequency and memory 
footprint, varying background load, transient network interrupts and so on, the service provision platform might 
face greatly slowdowns or even runtime crash. However, the default assignment, deployment, and management of containers 
does not \textit{fully} take the heterogeneity in both physical and virtualized nodes into consideration. Besides, 
the healing capability of Kubernetes is principally monitoring the status of containers, pods, and nodes and timely 
restarting the failures, which is not enough for high availability. Vayghan et al. find that in the specific test 
environment, when the pod failure happens, the outage time of the corresponding service could be dozens of seconds. 
When node failure happens, the outage time could be dozens of minutes \cite{Kubernetes-redundancy-1} \cite{Kubernetes-redundancy-2}. 
Therefore, with the vanilla version of Kubernetes, high availability might not be ensured, especially for the 
latency-critical cloud-native applications. Besides, one microservice could have several alternative 
execution solutions. For example, electronic payment, as a microservice of a composite service, can be executed by 
PayPal, WeChat Pay, and AliPay\footnote{Both Alipay and WeChat Pay are third-party mobile and online payment platforms, 
established in China.}. In this paper, let us call them \textit{candidates (of microservices)}. This status quo 
complicates the placement problem further. Because it is the instances of candidates that need to be placed, which 
greatly scales up the problem.

In order to solve the above problems, we propose a distributed redundant placement framework, 
i.e., Sample Average Approximation-based Redundancy Placement (SAA-RP), for a microservice-based applications 
with sequential combinatorial structure. For this application, if all of the 
candidates are placed on one edge site, network congestion is inevitable. Therefore, we adopt a distributed 
placement scheme, which is naturally suitable for the distributed edge. Redundancy is the core of SAA-RP, which 
allows that one candidate to be dispatched to multiple edge sites. By creating multiple candidate instances, 
it boosts a faster response to service requests. To be specific, it alleviates the risk of a long delay incurred 
when a candidate is assigned to only one edge site. With one candidate deployed on more than one 
edge site, requests from different end users at different locations can be balanced, so as to ensure the high availability 
of service and the robustness of the provision platform. Actually, performance of redundancy has been extensively 
studied under various system models and assumptions, such as the Redundancy-$d$ model, the $(n, k)$ system, and the 
$S\&X$ model \cite{Redundancy-model}. However, which kind of candidate requires redundancy and how many instances 
should be deployed cannot be decided if out of a concrete situation. Currently, the main strategy of job redundancy 
usually releases the resource occupancy after completion, which is not befitting for geographically distributed 
edge sites. This is because service requests are continuously generated from different end users. The destruction 
of candidate instances have to be created again, which will certainly lead to the delay in service responses. 
Besides, redundancy is not always a win and might be dangerous sometimes, since practical studies have shown that 
creating too many instances can lead to unacceptably high response times and even instability \cite{Redundancy-model}. 

As a result, we do not release the candidate instances but periodically update them based on the observations of 
service demand status during that period. Specifically, we derive expressions to decide each candidate should 
be dispatched with how many instances and which edge sites to place them. By collecting user requests for different 
service composition schemes, we model the distributed redundant placement as a stochastic discrete optimization 
problem and approximate the expected value through Monte Carlo sampling. During each sampling, we solve the deterministic 
problem based on an efficient evolutionary algorithm. Performance analysis and numerical results are provided to verify 
its practicability. Our main contributions are as follows.
\begin{enumerate}
  \item We model the distributed placement scenario at the edge for general microservice-based 
  chained applications and design a distributed redundant placement framework SAA-RP. SAA-RP can decide each candidate 
  should be dispatched with how many instances and which edge sites to placement them. It makes up with the shortcoming 
  of the default scheduler of Kubernetes, i.e. \texttt{kube-scheduler}\cite{Kube-scheduler}, when encountering the MEC.
  
  \item We take both the uncertainty of end users' service requests and the heterogeneity of edge sites into 
  consideration and formulate a stochastic optimization problem. Based on the long-term observation on end users' 
  service requests, we approximate the stochastic problem by sampling and averaging. 

  \item Simulations are conducted based on the real-world EUA Dataset \cite{EUA}. We also provide the performance analysis on 
  algorithm optimality and convergence rate. The numerical results verify the practicability and superiority 
  of our algorithm, compared with several typical benchmark policies. 
\end{enumerate}

The organization of this paper is as follows. Section \ref{sec3} demonstrates the motivation scenario. Section 
\ref{sec4} introduces the system model and formulates a stochastic optimization problem. The SAA-RP framework 
is proposed in Section \ref{sec5} and its performance analysis is conducted in Section \ref{sec6}. We show the 
simulation results in Section \ref{sec7}. In Section \ref{sec2}, we review related works on service placement at 
the edge and typical redundancy models. Section \ref{sec8} concludes this paper.

\section{Motivation Scenarios}\label{sec3}
\subsection{The Heterogeneous Network}\label{sec3.1}

\begin{figure}[!t]
  \centering
  \includegraphics[width=3.5in]{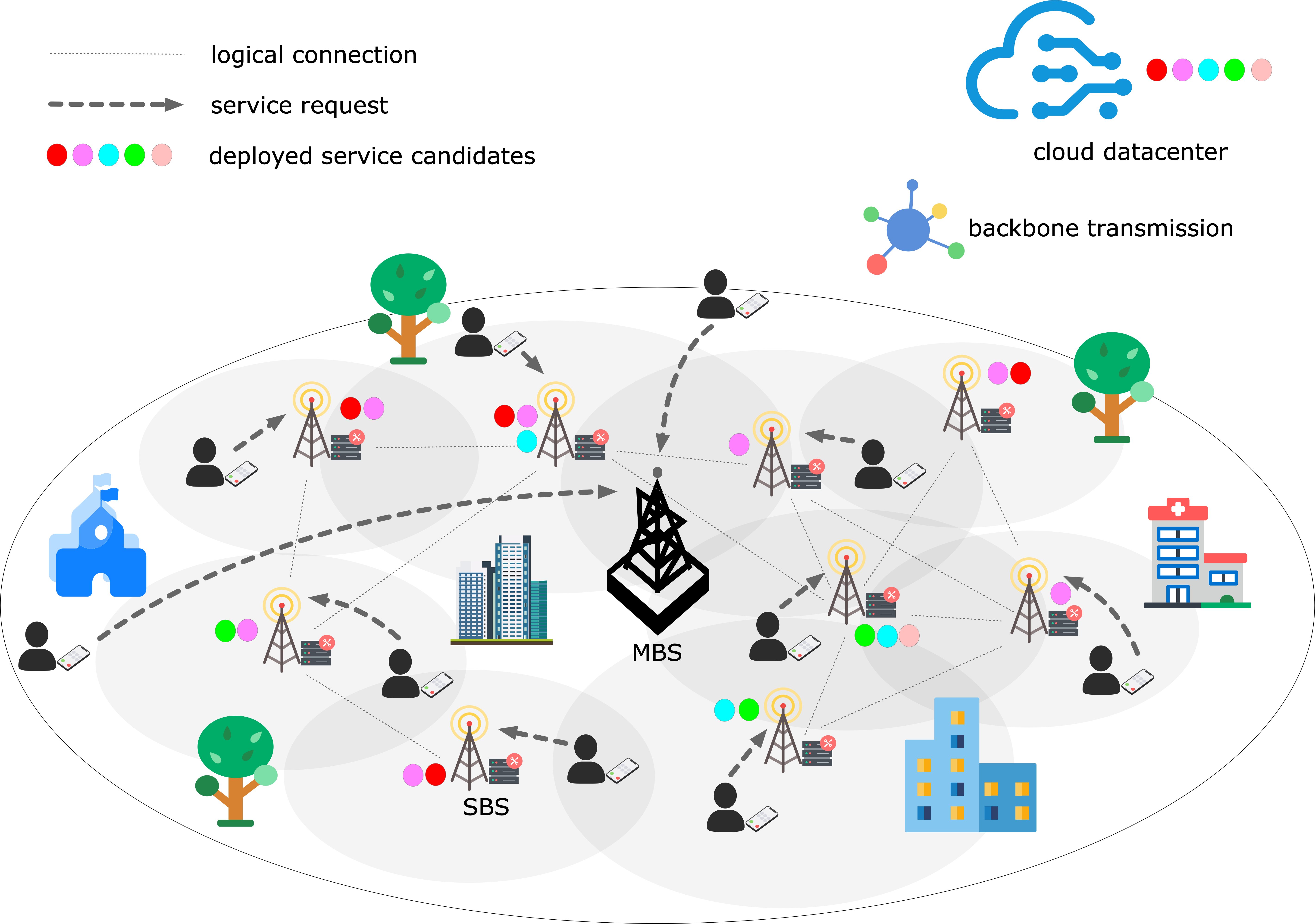}
  \caption{A typical scenario for a pre-5G HetNet.}
  \label{fig1}
\end{figure}

Let us consider a typical scenario for the pre-5G Heterogeneous Network (HetNet), which is the physical foundation of 
redundant service placement at the edge. As demonstrated in Fig. \ref{fig1}, for a given region, the 
wireless infrastructure of the access network can be \textit{simplified} into a Macro Base Station (MBS) and several 
Small-cell Base Stations (SBSs). The MBS is indispensable in any HetNet to provide ubiquitous coverage and support 
capacity, whose cell radius ranges from 8km to 30km. The SBSs, including femtocells, micro cells, and pico cells, are 
part of the network \textit{densification} for densely populated urban areas. Without loss of generality, WiFi access 
points, routers, and gateways are viewed as SBSs for simplification. Their cell radius ranges from 0.01km to 2km. The 
SBSs can be \textit{logically} interconnected to transfer signaling, broadcast message, and select 
routes. It might be too luxurious if all SBSs are fully interconnected, and not necessarily achievable if they are set 
up by different Mobile Telecom Carriers (MTCs)\footnote{Whether SBSs are logically connected comes down to their IP 
segments.}, but we reasonably assume that each SBS is mutually reachable to formulate \textit{an undirected connected 
graph}. This can be seen in Fig. \ref{fig1}. Each SBS has a corresponding small-scale data-center attached for the 
deployment of microservices and the allocating of resources. 

In this scenario, end users with their mobile devices can move arbitrarily within a certain range. For example, end 
users work within a building or rest at home. In this case, the connected SBS of each end user does not change.  

\subsection{Response Time of Microservices}\label{sec3.2}
A microservice-based application consists of multiple microservices. Each microservice can be executed by many 
available candidates. Take an arbitrary e-commerce application as an example. When we shop on a client browser, we 
firstly search the items we want, which can be realized by many site search APIs. Secondly, we add them to the cart 
and pay for them. The electronic payment can be accomplished by Alipay, WeChat Pay, or PayPal by invoking their 
APIs. After that, we can review and rate for those purchased items. In this example, each microservice is focused on single 
business capability. In addition, the considered application might have complex compositional structures and complex 
correlations between the fore-and-aft candidates because of \textit{bundle sales}. For example, when we are shopping 
on Taobao\footnote{Taobao is the world's biggest e-commerce website, established in China.}, only Alipay is supported 
for online payment. The application in the above example has a linear structure. As a beginning, this paper only cares 
about the \textit{sequentially} composed application. In practice, a general directed acyclic graph (DAG) can be 
decomposed into several linear chains by applying Flow Decomposition Theorem (located in Chapter 03) \cite{network_flows}. 
We leave the extension to future work.

The pre-5G HetNet allows SBSs to share a mobile service provision platform, where user configurations and 
contextual information can be uniformly managed. As we have mentioned before, the unified platform can be 
implemented by Kubernetes. In our scenario, each mobile device sends its service request to the nearest SBS 
for the strongest signal of the established link. However, if there are no SBSs accessible, the request 
has to be responded by the MBS and processed by cloud data-centers. All the possibilities of the response status of 
the first microservice is discussed below.
\begin{enumerate}
  \item The requested candidate is deployed at the chosen SBS. It will be processed by this SBS instantly.
  \item The requested candidate is not deployed at the nearest SBS but accessible on other SBSs, which leads 
  to multi-hop transfers between the SBSs until the request is responded by another SBS. That is, the request 
  will \textit{route through} the HetNet until it is responded by an SBS who deploys the required candidate.
  \item The requested candidate is not deployed on any SBSs in the HetNet. It can only be processed 
  by cloud through backbone transmission.
\end{enumerate}

For the subsequent microservices, the response status also faces many possibilities: 
\begin{enumerate}
  \item The previous candidate is processed by an SBS. Under this circumstance, for candidate of this microservice, 
  if its instance can be found in the HetNet, multi-hop transfer is required. Otherwise, it has to be 
  processed by cloud.
  \item The previous candidate is processed by cloud. Under this circumstance, the candidates of subsequent microservices 
  should always be responded by cloud without unnecessary backhaul.
\end{enumerate}
Our job is to find an optimal redundant placement policy with the trade-offs between resource occupation 
and response time considered. We should know which candidates who might as well be redundant and where to 
deploy them.

\subsection{A Working Example}\label{sec3.3}
This subsection describes a small-scale working example. 

\begin{figure}[!t]
  \centering
  \includegraphics[width=2.5in]{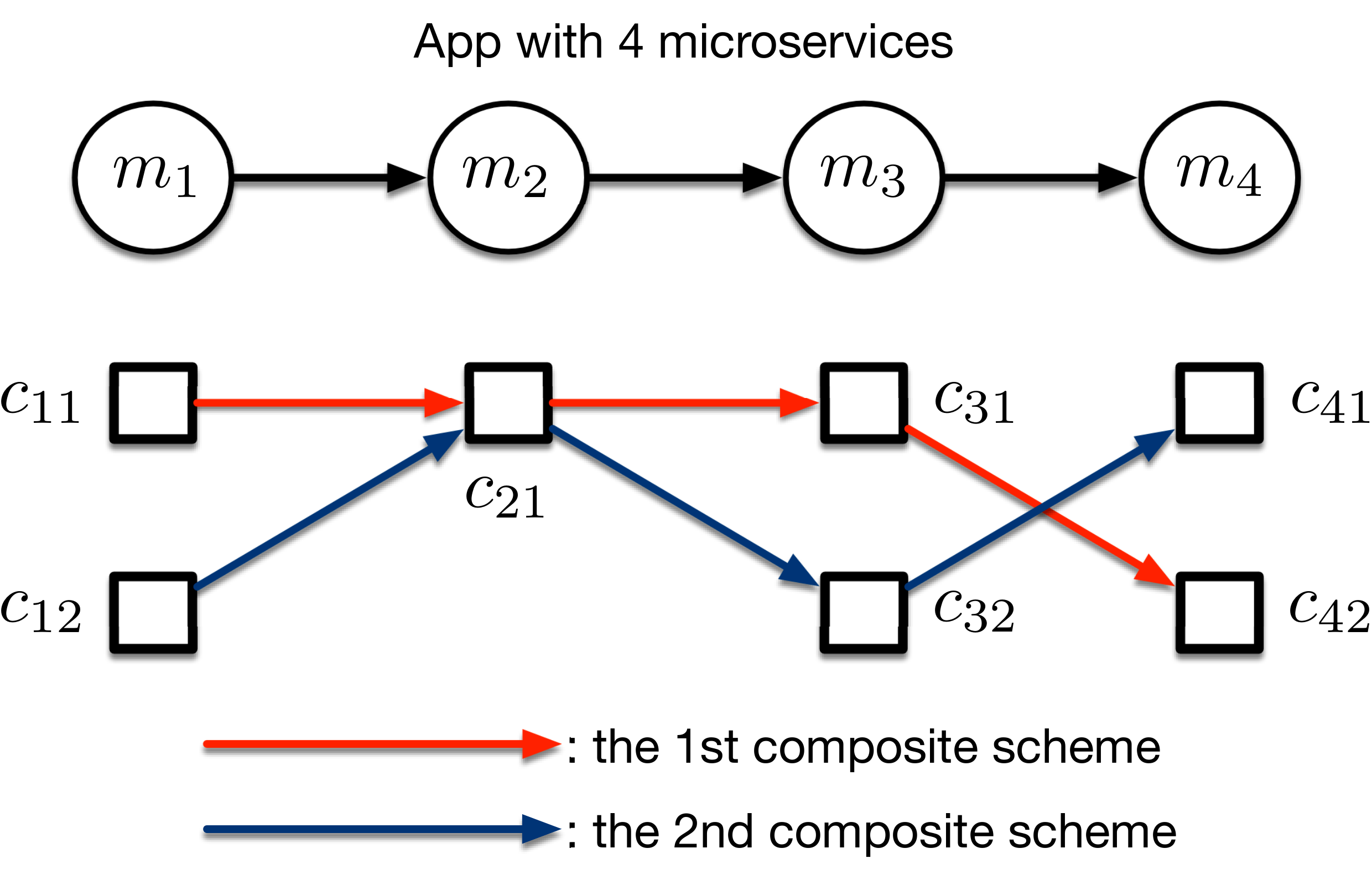}
  \caption{Two service composition schemes for a 4-microservice app.}
  \label{fig_add1}
\end{figure}

\begin{figure}[!t]
  \centering
  \includegraphics[width=2.65in]{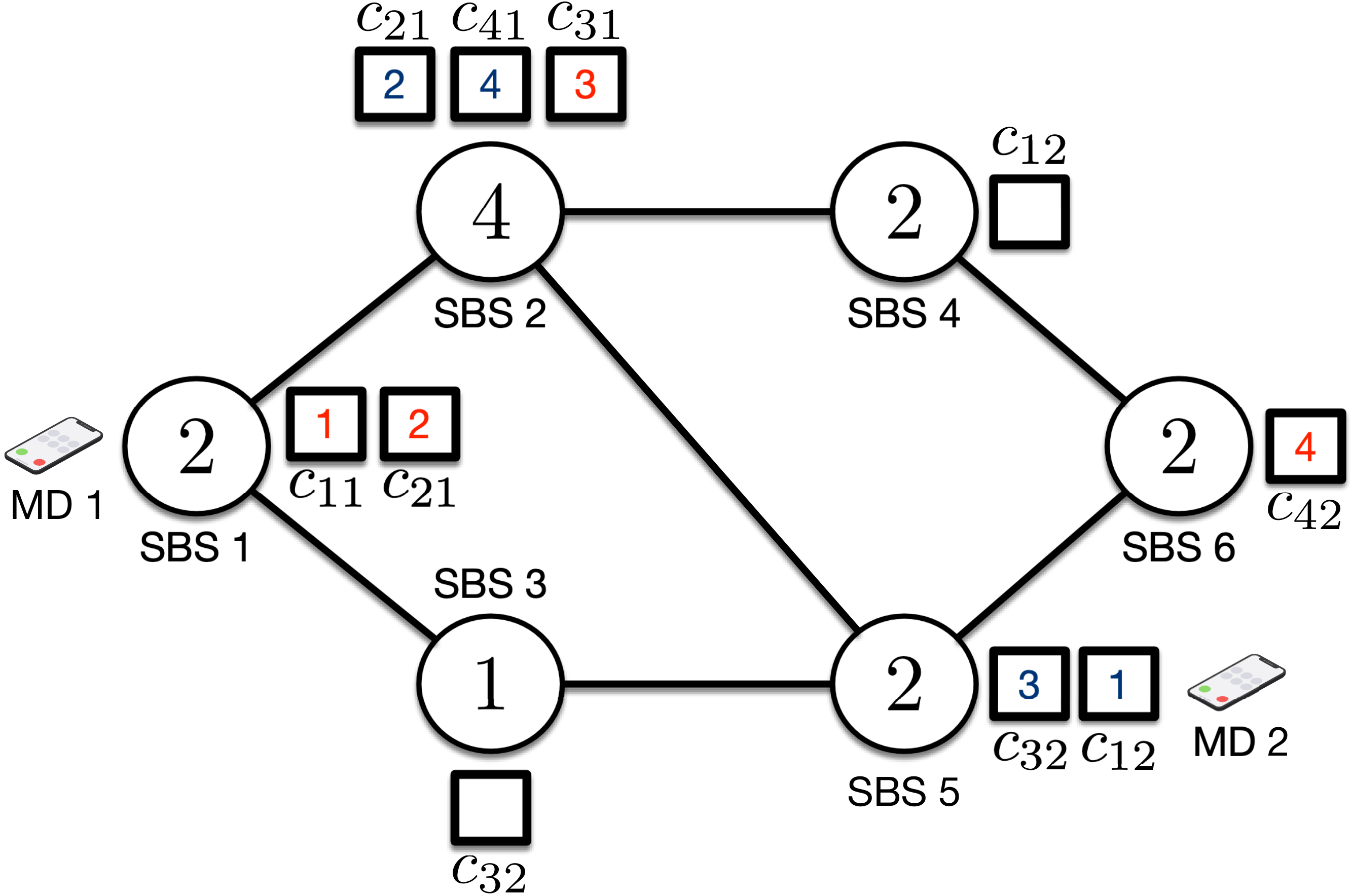}
  \caption{The placement of each candidates on the HetNet.}
  \label{fig_add2}
\end{figure}

\textit{\textbf{Microservices and Candidates:}} Fig. \ref{fig_add1} demonstrates a chained application constitutive of 
four microservices. Each microservice has 2, 1, 2, and 2 candidates, respectively. The first service composition 
scheme is $c_{11} \to c_{21} \to c_{31} \to c_{42}$, and the second service composition scheme 
is $c_{12} \to c_{21} \to c_{32} \to c_{41}$. In practice, the composition scheme is decided by the daily usage habits 
of end users. It might be strongly biased. Besides, because of bundle sales, part of the composition might 
be fixed. We will describe the composition in terms of a joint probability distribution in Subsection \ref{sec4.1}. 

\textit{\textbf{Service Placement of Instances:}} In Fig. \ref{fig_add2}, the undirected connected graph consists of 
6 SBSs. The number tagged inside each SBS is its maximum number of placeable candidates. For example, SBS2 
can be placed at most 4 candidates. This constraint exists because the edge sites have very limited computation and 
storage resources, compared with cloud data-centers. The squares beside each SBS are the deployed candidates. For example, 
SBS1 deploys two candidates, $c_{11}$ and $c_{21}$. Notice that because of the redundancy mechanism, 
the same candidate can be deployed on multiple SBSs. For example, $c_{21}$ is dispatched to both SBS1 and SBS2. 

Fig. \ref{fig_add2} also demonstrates two 
mobile devices, MD1 and MD2, which are located beside SBS1 and SBS5, respectively. It means that the SBSs 
closest to MD1 and MD2 are SBS1 and SBS5, respectively. As we have already mentioned in Subsection \ref{sec3.2}, 
the service request of each mobile device is responded by the nearest SBS. Thus, for MD1 and MD2, SBS1 and 
SBS5 are the corresponding SBS for responding, respectively. 

\textit{\textbf{Response Time Calculation:}} We assume that the service composition scheme of MD1 is the red one 
in Fig. \ref{fig_add1}, and MD2's is the blue one. The number tagged inside each candidate is the executing sequence. 
Let us take a closer look at MD1. 
\begin{enumerate}
  \item $c_{11}$: Because $c_{11}$ is deployed on SBS1, the response time of $c_{11}$ is equal to the sum of the 
  expenditure of time on wireless access between MD1 and SBS1 and the processing time of $c_{11}$ on SBS1. 

  \item $c_{21}$: Because $c_{21}$ can also be found on SBS1, the expenditure time is zero. The response time 
  of $c_{21}$ consists of only the processing time of $c_{21}$ on SBS1. 

  \item $c_{31}$: $c_{31}$ can only be found on SBS2, thus the expenditure time of it is equal to the routing time 
  from SBS1 to SBS2. In this paper, we assume that the routing between two nodes always selects the nearest path in 
  the undirected graph. Thus, only one hop is required (SBS1 $\to$ SBS2). Thus, the response time of $c_{31}$ consists 
  of the routing time from SBS1 to SBS2 and the processing time of $c_{31}$ on SBS2.

  \item $c_{42}$: $c_{42}$ can only be found on SBS6. Thus, the expenditure requires 2 hops (SBS2 $\to$ SBS4 $\to$ 
  SBS6 or SBS2 $\to$  SBS5 $\to$ SBS6). Finally, the output of $c_{42}$ need to be transferred back to MD1 via SBS1. The nearest path from SBS6 
  to SBS1 requires 3 hops. There are three alternatives: $(i)$ SBS6 $\to$ SBS4 $\to$ SBS2 $\to$ SBS1 or $(ii)$ 
  SBS6 $\to$ SBS5 $\to$ SBS2 $\to$ SBS1 or $(iii)$ SBS6 $\to$ SBS5 $\to$ SBS3 $\to$ SBS1. Thus, the response time 
  of $c_{42}$ consists of the routing time from SBS2 to SBS6, the processing time of $c_{42}$ on SBS6, the routing time from 
  SBS6 to SBS1, and the wireless transmission time from SBS1 to MD1.
\end{enumerate}
The response time of the $1^{st}$ composition scheme is the sum of the response time of $c_{11}$, $c_{21}$, $c_{31}$, and $c_{42}$. 
The same procedure applies to MD2. In addition, there are two unexpected cases need to be addressed. The 
first one is that if a mobile device is covered by no SBS, the response should be made by the MBS and all the 
microservices are processed by cloud. The second one is that if a required candidate is not deployed on any SBS, 
then a communication link from the SBS processing the last candidate to the cloud should be established. This 
candidate and all the candidates of the rest microservices will be processed on cloud. The response time is 
calculated in a different way for these cases. Nevertheless, all the contingencies are taken into account in 
our system model, elaborated in Section \ref{sec4}.

Obviously, a better service placement policy can lead to less time spent. Our job is to find a service placement policy 
to minimize the response time of all mobile devices. What need to determine are not only how many instances required 
for each candidate, but also which edge sites to place them. The next section will demonstrate the rigorous formulation of system 
model.


\section{System Model}\label{sec4}
\newcommand{\tabincell}[2]{\begin{tabular}{@{}#1@{}}#2\end{tabular}}
  \begin{table}
    \centering
    \caption{Summary of key notations.}
    \begin{tabular}{ll}
      \\[-2mm]\hline\hline\\[-2mm]
      {\bf Notation}&\qquad \qquad {\bf Description}\\[+1mm]
      \hline\\[-2mm]
      $M$ & \tabincell{l}{the number of SBSs, $M = |\mathcal{M}|$}\\[+0.7mm]
      $N$ & \tabincell{l}{the number of mobile devices, $N = |\mathcal{N}|$}\\[+0.7mm]
      $Q$ & \tabincell{l}{the number of microservices in the application,\\\quad $Q = |\mathcal{Q}|$}\\[+0.7mm]
      $t_q, q \in \mathcal{Q}$ & \tabincell{l}{the $q$th microservice in the application}\\[+0.7mm]
      $\mathcal{M}_i$ & \tabincell{l}{the set of SBSs that can be connected by the\\\quad $i$th mobile device}\\[+0.7mm]
      $\mathcal{N}_j$ & \tabincell{l}{the set of mobile devices that can be\\\quad connected by the $j$th SBS}\\[+0.7mm]
      $C_q$ & \tabincell{l}{the number of candidates of the $q$th\\\quad microservice, $C_q = |\mathcal{C}_q|, q \in \mathcal{Q}$}\\[+0.7mm]
      $s^c_q, c \in \mathcal{C}_q$ & \tabincell{l}{the $c$th candidate of the $q$th microservice}\\[+0.7mm]
      $\mathcal{D}(s^c_q)$ & \tabincell{l}{the set of SBSs on which $s^c_q$ is deployed}\\[+0.7mm]
      $E(s^c_q)$ &	\tabincell{l}{the random event that for microservice $t_q$, the $c$th\\\quad candidate is selected for execution}\\[+0.7mm]
      $j^\star_i$ & \tabincell{l}{the nearest SBS to the $i$th mobile device}\\[+0.7mm]
      $j_p(s^c_q)$ & \tabincell{l}{the SBS who actually processes the candidate $s^c_q$}\\[+0.7mm]
      $d(i,j)$ & \tabincell{l}{the Euclidean distance between the $i$th mobile\\\quad device and the $j$th SBS}\\[+0.7mm]
      $\tau_{in}(s^c_q)$ & \tabincell{l}{the data uplink transmission time of the\\\quad candidate $s^c_q$}\\[+0.7mm]
      $\tau_{exe}(j_p(s^c_q))$ & \tabincell{l}{the execution time of the candidate $s^c_q$ on\\\quad the $j_p$th SBS}\\[+0.7mm]
      $\zeta(j_1,j_2)$ & \tabincell{l}{the hop-count between the $j_1$th SBS\\\quad and the $j_2$th SBS}\\[+0.7mm]
      $b_j$ & \tabincell{l}{the maximum number of microservice instances \\\quad can be deployed on the $j$th SBS}\\[+2mm]
      \hline\hline
    \end{tabular}
    \label{tab1}
\end{table}
The HetNet consists of $N$ mobile devices, indexed by $\mathcal{N} \triangleq 
\{1, ..., i, ..., N\}$, $M$ SBSs, indexed by $\mathcal{M} \triangleq \{1, ..., j, ..., M\}$, and one MBS, 
indexed by $0$. Considering that each mobile device might be covered by many SBSs, let us use $\mathcal{M}_i$ 
to denote the set of SBSs whose wireless signal covers the $i$th mobile devices. Correspondingly, $\mathcal{N}_j$ 
denotes the set of mobile devices that are covered by the $j$th SBS. Notice that the service request 
from a mobile device is responded by its nearest available SBS, otherwise the MBS. The MBS here is to provide 
ubiquitous signal coverage and is always connectable to each mobile device. Both the SBSs and the MBS are 
connected to the backbone. Table \ref{tab1} lists key notations in this paper. 

\subsection{Describing the Correlated Microservices}\label{sec4.1}
We consider an application with $Q$ sequential composite microservices $\langle t_1, ..., t_Q \rangle$, 
indexed by $\mathcal{Q}$. $\forall q \in \mathcal{Q}$, microservice $t_q$ has $C_q$ candidates 
$\{ s^1_q, ..., s^{c}_q, ..., s^{C_q}_q \}$, indexed by $\mathcal{C}_q$. Let us use 
$\mathcal{D}(s^c_q) \subseteq \mathcal{M}$ to denote the set of SBSs on which $s^c_q$ is deployed. Our 
redundancy mechanism allows that $|\mathcal{D}(s^c_q)| > 1$, which means one candidate instance 
could be dispatched to more than one SBS. Besides, let us use $E(s^c_q), c \in \mathcal{C}_q$ to represent that 
for microservice $t_q$, the $c$th candidate is selected for execution. $E(s^c_q)$ 
can be viewed as a random event with an unknown distribution. Further, $\mathbb{P}(E(s^c_q))$ denote the probability that 
$E(s^c_q)$ happens. Thus, for each mobile device, its selected candidates can be described 
as a $Q$-tuple: 
\begin{equation} \label{prob} \tag{1}
  (\vec{s}) \triangleq
  \langle E(s^{c_1}_1), ..., E(s^{c_Q}_Q) \rangle, 
\end{equation}
where $q \in \mathcal{Q}, c_q \in \mathcal{C}_q$. 

The sequential composite application might have correlations between the fore-and-aft candidates. The definition below 
gives a rigorous mathematical description.
\begin{definition}
	\textbf{Correlations of Composite Service}
  $\forall q \in \mathcal{Q}\backslash\{1\}$, $c_1 \in \mathcal{C}_{q-1}$, and $c_2 \in \mathcal{C}_q$, candidate 
  $s^{c_2}_q$ and $s^{c_1}_{q-1}$ are correlated \textit{iff} $\mathbb{P}(E(s^{c_2}_q) | E(s^{c_1}_{q-1})) \equiv 1$, 
  and $\forall c_2' \in \mathcal{C}_q \backslash \{c_2\}$, $\mathbb{P}(E(s^{c_2'}_q) | E(s^{c_1}_{q-1})) \equiv 0$.
\end{definition}
With the above definition, the probability $\mathbb{P} (E (\vec{s}))$ can be calculated by
\begin{align} \label{scheme}
	&\qquad \qquad \qquad \mathbb{P} (E (\vec{s})) = \notag\\
	&\mathbb{P}(E(s^{c_1}_1)) \cdot \prod_{q=2}^Q \mathbb{P}(E(s^{c_q}_q) | E(s^{c_{q-1}}_{q-1})). \tag{2}
\end{align}

\subsection{Calculating the Respone Time}
The response time of one candidate consists of data uplink transmission time, service execution time, and 
data downlink transmission time. The data uploaded is mainly encoded service requests and configurations while 
the output is mainly the feedback on successful service execution or a request to invoke the next candidate. 
If all requests are responded within the access network, most of the time is spent on routing with multi-hops between SBSs. 
Notice that except the last one, each candidate's data uplink transmission time is equal to the data downlink 
transmission time of the candidate of its previous microservice. 

\newcounter{mytempeqncnt}
\begin{figure*}[!t]
  \normalsize
\setcounter{mytempeqncnt}{\value{equation}}
\setcounter{equation}{2}
\begin{flalign}
	&j_p(s^{c_1}_1(i)) =
  \left\{
    \begin{array}{ll}
      \textrm{cloud},  & \mathcal{M}_i = \varnothing \textrm{ or } \mathcal{D}(s^{c_1}_1(i)) = \varnothing; \\
      j^\star_i,  & \mathcal{D}(s^{c_1}_1(i)) \neq \varnothing, j^\star_i \in \mathcal{D}(s^{c_1}_1(i)); \\
      \argmin_{j^\bullet \in \mathcal{D}(s^{c_1}_1(i))} \zeta(j^\star_i, j^\bullet),  & \textrm{otherwise} \\
		\end{array}
	\right.&
	\label{j_p}
\end{flalign}

\begin{flalign}
	&\tau_{in}(s^{c_1}_1(i)) =
  \left\{
    \begin{array}{ll}
      \alpha \cdot d(i, 0) + \tau_{b},  & \mathcal{M}_i = \varnothing; \\
      \alpha \cdot d(i, j^\star_i) + \tau_{b},  & \mathcal{M}_i \neq \varnothing, \mathcal{D}(s^{c_1}_1(i)) = \varnothing; \\
			\alpha \cdot d(i, j^\star_i), & \mathcal{M}_i \neq \varnothing, j^\star_i \in \mathcal{D}(s^{c_1}_1(i)); \\
      \alpha \cdot d(i, j^\star_i) + \beta \cdot \min_{j^\bullet \in \mathcal{D}(s^{c_1}_1(i))} \zeta(j^\star_i, j^\bullet),  & \textrm{otherwise} \\
		\end{array}
	\right.&
	\label{s1_time}
\end{flalign}

\begin{flalign}
	&j_p(s^{c_q}_q(i)) =
  \left\{
    \begin{array}{ll}
      \textrm{cloud},  & \mathcal{M}_i = \varnothing \textrm{ or } \mathcal{D}(s^{c_q}_q(i)) = \varnothing; \\
      j_p(s^{c_{q-1}}_{q-1}(i)),  & \mathcal{D}(s^{c_q}_q(i)) \neq \varnothing, j_p(s^{c_{q-1}}_{q-1}(i)) \in \mathcal{D}(s^{c_q}_q(i)); \\
      \argmin_{j^\bullet \in \mathcal{D}(s^{c_q}_q(i))} \zeta(j_p(s^{c_{q-1}}_{q-1}(i)), j^\bullet),  & \textrm{otherwise} \\
		\end{array}
	\right.&
	\label{j_p_2}
\end{flalign}

\begin{flalign}
	&\tau_{in}(s^{c_q}_q(i)) = 
	\left\{
    \begin{array}{ll}
      0, &j_p(s^{c_{q-1}}_{q-1}(i)) \in \mathcal{D}(s^{c_q}_q(i)) \textrm{ or } j_p(s^{c_{q-1}}_{q-1}(i)) = \textrm{cloud}; \\
      \tau_b, &j_p(s^{c_{q-1}}_{q-1}(i)) \neq 0, \mathcal{D}(s^{c_q}_q(i)) = \varnothing; \\
      \beta \cdot \min_{j^\bullet \in \mathcal{D}(s^{c_q}_q(i))} \zeta(j_p(s^{c_{q-1}}_{q-1}(i))), j^\bullet), &\textrm{otherwise} \\
		\end{array}
	\right.&
	\label{s_q_time}
\end{flalign}

\begin{flalign}
	&\tau_{out}(s^{c_Q}_Q(i)) = 
	\left\{
    \begin{array}{ll}
      \tau_b + \alpha \cdot d(i, 0), &j_p(s^{c_Q}_Q(i)) = \textrm{cloud}; \\
      \beta \cdot \zeta(j_p(s^{c_Q}_Q(i)), j^\star_i) + \alpha \cdot d(i, j^\star_i), &\textrm{otherwise} \\
		\end{array}
	\right.&
	\label{s_q_time_out}
\end{flalign}
\setcounter{equation}{\value{mytempeqncnt}}
\hrulefill
\vspace*{4pt}
\end{figure*}

\subsubsection{For the Initial Candidate}
For the $i$th mobile device and its selected candidate $s^{c_1}_1(i)$ of the initial microservice $t_1$, where 
$c_1 \in \mathcal{C}_1$, (\textbf{I}) if the $i$th mobile device is not covered by any SBSs around, i.e., $\mathcal{M}_i = \varnothing$, the request has 
to be responded by the MBS and processed by cloud data-center. (\textbf{II}) If $\mathcal{M}_i \neq \varnothing$, 
as we have mentioned before, the nearest SBS $j^\star_i \in \mathcal{M}_i$ is chosen and connected. 
Under this circumstance, a classified discussion is required. 
\begin{enumerate}
  \item If the candidate $s^{c_1}_1(i)$ is not deployed on any SBSs from $\mathcal{M}$, i.e. 
  $\mathcal{D}(s^{c_1}_1(i)) = \varnothing$, the request still has to be forwarded to cloud data-center through 
  backbone transmission.
  \item If $\mathcal{D}(s^{c_1}_1(i)) \neq \varnothing$ and 
  $j^\star_i \in \mathcal{D}(s^{c_1}_1(i))$, the request can be directly processed by SBS 
  $j^\star_i$ without any hops. 
  \item If $\mathcal{D}(s^{c_1}_1(i)) \neq \varnothing$ and 
  $j^\star_i \notin \mathcal{D}(s^{c_1}_1(i))$, the request has to be responded by 
  $j^\star_i$ and processed by another SBS from $\mathcal{D}(s^{c_1}_1(i))$. 
\end{enumerate}
We use $j_p(s)$ to denote the SBS who actually processes $s$. Remember that the 
routing between two nodes always selects the nearest path. Thus, for $q = 1$, $j_p(s^{c_1}_1(i))$ can be obatined 
by \eqref{j_p}, where $\zeta(j_1,j_2)$ is the shortest number of hops from node $j_1$ to node $j_2$.

In this paragragh, we calculate the response time of $s^{c_1}_1(i)$. We use $d(i,j)$ to denote the reciprocal of 
the bandwidth between $i$ and $j$. Obviously, the expenditure of time on wireless access is inversely proportional to 
the bandwidth of the link. Besides, the expenditure of time on routing is directly proportional to the number of 
hops between the source node and destination node. As a result, the data uplink transmission time $\tau_{in}(s^{c_1}_1(i))$ 
is summarized in \eqref{s1_time}, where $\tau_{b}$ is the time on backbone transmission, $\alpha$ is size of input data 
stream from each mobile device to the initial candidate (measured in bits), and $\beta$ is a positive constant 
representing the rate of wired link between SBSs. We use $\tau_{exe}(j_p(s^{c_1}_1(i)))$ to denote the microservice execution time 
on the SBS $j_p$ for the candidate $s^{c_1}_1(i)$. The data downlink transmission time is the same as the uplink 
time of the next microservice, which is discussed hereinafter.

\subsubsection{For the Intermediate Candidates}
For the $i$th mobile device and its selected candidate $s^{c_q}_q(i)$ of microservice $t_q$, where 
$q \in \mathcal{Q} \backslash \{1, Q\}$, $c_q \in \mathcal{C}_q$, the analysis of its data uplink transmission 
time is correlated with $j_p(s^{c_{q-1}}_{q-1}(i))$, i.e. the SBS who processes $s^{c_{q-1}}_{q-1}(i)$. 
$\forall q \in \mathcal{Q} \backslash \{1\}$, the calculation of $j_p(s^{c_q}_q(i))$ is 
summarized in \eqref{j_p_2}. This formula is closely related to \eqref{j_p}. 

In this paragragh, we calculate the response time of $s_q^{c_q}(i)$.
(\textbf{I}) If $j_p(s^{c_{q-1}}_{q-1}(i)) \in \mathcal{D}(s^{c_q}_q(i))$, then the data downlink transmission 
time of previous candidate $s^{c_{q-1}}_{q-1}(i)$, which is also the data uplink transmission time of current candidate  
$s^{c_q}_q(i)$, is zero. That is, $\tau_{out}(s^{c_{q-1}}_{q-1}(i)) = \tau_{in}(s^{c_q}_q(i)) = 0$. It is because both 
$s^{c_{q-1}}_{q-1}(i)$ and $s^{c_q}_q(i)$ are deployed on the SBS $j_p(s^{c_{q-1}}_{q-1}(i))$, which leads to 
the number of hops being zero. (\textbf{II}) If $j_p(s^{c_{q-1}}_{q-1}(i)) \notin \mathcal{D}(s^{c_q}_q(i))$, 
a classified discussion has to be launched. 
\begin{enumerate}
  \item If $j_p(s^{c_{q-1}}_{q-1}(i)) = \textrm{cloud}$, which means the request of $t_{q-1}$ from 
  the $i$th mobile device is responded by cloud data-center. In this case, the invocation for $t_q$ can be directly processed 
  by cloud without backhaul. Thus, the data uplink transmission time of $t_q$ is zero, too. 
  \item If $j_p(s^{c_{q-1}}_{q-1}(i)) \neq 0$ and $\mathcal{D}(s^{c_q}_q(i)) = \varnothing$, which means 
  $s^{c_q}_q(i)$ is not deployed on any SBSs in the HetNet. As a result, the invocation for $t_q$ has to be 
  responded by cloud data-center through backbone transmission. 
  \item If $j_p(s^{c_{q-1}}_{q-1}(i)) \neq 0$, $\mathcal{D}(s^{c_q}_q(i)) \neq \varnothing$ but 
  $j_p(s^{c_{q-1}}_{q-1}(i)) \notin \mathcal{D}(s^{c_q}_q(i))$, which means both $s^{c_{q-1}}_{q-1}(i)$ 
  and $s^{c_q}_q(i)$ are processed by the SBSs in the HetNet but not the same one. In this case, we can calculate the 
  data uplink transmission time by finding the shortest path from $j_p(s^{c_{q-1}}_{q-1}(i))$ to a SBS in 
  $\mathcal{D}(s^{c_q}_q(i))$.
\end{enumerate}
The above analysis is summarized in \eqref{s_q_time}.

\subsubsection{For the Last Candidate}
For the $i$th mobile device and its selected candidate $s^{c_Q}_Q(i)$ of the last microservice $t_Q$, where $c_Q \in \mathcal{C}_Q$, the data 
uplink transmission time $\tau_{in}(s^{c_Q}_Q(i))$ is also calculated by \eqref{s_q_time}, with every $q$ 
replaced by $Q$. However, for the data downlink transmission time $\tau_{out}(s^{c_Q}_Q(i))$, a classified 
discussion is required: (\textbf{I}) If $j_p(s^{c_Q}_Q(i)) = 0$, which means the chosen candidate of the last 
microservice is processed by cloud data-center, the processed result need to be returned from cloud to the $i$th 
mobile device through backhaul transmission\footnote{We assume that the backhaul can only 
be transferred through the MBS.}. (\textbf{II}) If $j_p(s^{c_Q}_Q(i)) \neq 0$, which means the chosen candidate of 
the last microservice is processed by a SBS in the HetNet. In this case, the result should be delivered to the $i$th 
mobile device via $j_p(s^{c_Q}_Q(i))$ and $j^\star_i$\footnote{$j_p(s^{c_Q}_Q(i))$ 
and $j^\star_i$ could be the same one. In this case, the number of hops between them is zero.}.
\eqref{s_q_time_out} summarizes the calculation of $\tau_{out}(s^{c_Q}_Q(i))$.

Based on the above analysis, the response time of the $i$th mobile device is 
\begin{align}
  \tau(E (\vec{s}(i)))
  &= \sum_{q=1}^Q \Big( \tau_{in}(s^{c_q}_q(i)) + \tau_{exe}\big(j_p(s^{c_q}_q(i))\big) \Big) \notag\\
  &+ \tau_{out}(s^{c_Q}_Q(i)). \tag{8}
  \label{total_time}
\end{align}

So far, the system model has been elaborated. The assumptions in this paper are summarized as follows.

1) We assume that the edge sites can form an undirected connected graph. In MEC, this assumption is rational 
and frequently-used, especially for the research on Network Slicing \cite{network_slicing}. 

2) We assume that it is the nearest SBS that responses to the initial microservice of a mobile device. This 
assumption is naive and widely-used because it leads to the minimal first-step communication cost. 

3) We only consider the composite application with linear structure. As we have mentioned, a general DAG 
can be decomposed into several linear chains, thus we leave the extension to future work. 

4) We assume that the expenditure of time on routing is directly proportional to the number of hops. In 
MEC, the HetNet is a given region whose range is within tens of kilometers. The transmission time is mainly 
spent on routing and transit. Thus, this assumption is rational. 

5) We assume that the backhaul can only be transferred through the MBS. There are multiple alternatives 
for backhaul in 5G communications. We make the assumption to simplify the problem formulation for the chosen 
candidate of the last microservice. 

\subsection{Problem Formulation}
Our job is to find an optimal redundant placement policy to minimize the overall latency under the limited 
capability of SBSs. The heterogeneity of edge sites is directly embodied in the number of \textit{can-be-deployed} 
candidates. Let us use $b_j$ to denote this number for the $j$th SBS. In the heterogenous edge, $b_j$ could vary 
considerably. 
The following constraint should be satisfied: 
\begin{align}
  \sum_{q \in \mathcal{Q}} \sum_{c \in \mathcal{C}_q} 
  \mathds{1}\{j \in \mathcal{D}(s^{c_q}_q) \} \leq b_j, \forall j \in \mathcal{M}, \tag{9}
  \label{cons_1}
\end{align}
where $\mathds{1}\{\cdot\}$ is the indicator function. 
Finally, the optimal placement problem can be formulated as: 
\begin{align}
	\mathcal{P}_1: &\min_{\mathcal{D}(s^{c_q}_q)} \sum_{i=1}^N \tau(E (\vec{s}(i))) \notag\\
  s.t. & \qquad \quad \eqref{cons_1} \notag,
\end{align}
where the decision variables are $\mathcal{D}(s^{c_q}_q(i)), \forall q \in \mathcal{Q}, c \in \mathcal{C}_q$, 
and the optimization goal is the sum of response time of all mobile devices. 

\section{Algorithm Design}\label{sec5}
In this section, we elaborate our algorithm for $\mathcal{P}_1$. Firstly, we recode the decision variables as 
$\vec{x}$ to \textit{shrink} the size of feasible region. Based on that, we propose the SAA-RP framework. It 
includes a subroutine, named Genetic Algorithm-based Server Selection (GASS) algorithm. The details are presented 
as follows. 

\subsection{Variable Recoding}
Let us use $\vec{W}(i) \triangleq (\texttt{canIdx}(t_1), ..., \texttt{canIdx}(t_Q))$ to denote the random vector on 
the chosen service composition scheme of the $i$th mobile device, where $\texttt{canIdx}(t_q)$ returns the index 
of the chosen candidate of the $q$th microservice. $\vec{W}(i)$ and $E (\vec{s}(i))$ describe the 
same thing from different perspectives. However, the former is more concise. Then, we use $\vec{W} \triangleq (\vec{W}(1), ..., \vec{W}(N))$ 
to denote the global random vector by taking all mobile devices into account. 
Let us use $\vec{x} \triangleq [\vec{x}(b_1), ..., \vec{x}(b_M)]^\top$ to recode the global \textit{deploy-or-not} vector. 
$\forall j \in \mathcal{M}, \vec{x}(b_j)$ is a deployment vector for SBS $j$, whose length is $b_j$. By doing this, the constraint 
\eqref{cons_1} can be removed because it is reflected in how $\vec{x}$ encodes. $\forall j \in \mathcal{M}$, each element of 
$\vec{x}(b_j)$ is chosen from $\{0, 1, ..., \sum_{q=1}C_q\}$, i.e. the global index of each candidate. That is, any candidate 
can appear in any number of SBSs in the HetNet. Thus, the redundancy mechanism is also reflected in how $\vec{x}$ encodes.
$\forall j \in \mathcal{M}$, $\vec{x}(b_j) = \vec{0}$ means that the $j$th SBS does not deploy any candidate. 

$\vec{x}$ is a new encoding of $\mathcal{D}(s^{c_q}_q(i))$. As a result, we can reconstitute 
$\tau(E (\vec{s}(i)))$ as $\tau(\vec{x}, \vec{W}(i))$. As such, the optimization goal can be written as
\begin{align}
  g(\vec{x}) \triangleq \mathbb{E}[G(\vec{x}, \vec{W})] = \mathbb{E}[\sum_{i=1}^N \tau(\vec{x}, \vec{W}(i))], \tag{10}
  \label{true_goal}
\end{align}
and the optimal placement problem is 
\begin{align}
	\mathcal{P}_2: &\min_{\vec{x} \in \mathcal{X}} g(\vec{x}). \notag
\end{align}
$\mathcal{P}_2$ is a stochastic discrete optimization problem with independent variable $\vec{x}$, where $\mathcal{X}$ 
is the feasible region. $\mathcal{X}$, although finite, is very large. Therefore, 
enumeration approach is inadvisable. Besides, the problem has an uncertain random vector $\vec{W}$ with probability 
distribution $\mathbb{P} (E (\vec{s}(i)))$. 

\subsection{The SAA-RP Framework}
Let us take a closer look at $\mathcal{P}_2$. Firstly, the random vector $\vec{W}$ is \textit{exogenous} because the decision 
on $\vec{x}$ does not affect the distribution of $\vec{W}$. Secondly, for a given $\vec{W}$, $G(\vec{x}, \vec{W})$ 
can be easily evaluated for any $\vec{x}$. Thus, the observation of $G(\vec{x}, \vec{W})$ is \textit{constructive}. As a result, 
we can apply the Sample Average Approximation (SAA) approach to $\mathcal{P}_1$ \cite{SAA} to handle with the uncertainty. 

SAA is a classic Monte Carlo simulation-based method. In the following section, we elaborate how 
we apply the SAA method to $\mathcal{P}_2$. Formally, we define the SAA problem $\mathcal{P}_3$. Let 
$\vec{W}^1, \vec{W}^2, ..., \vec{W}^R$ be an independently and identically distributed (i.i.d.) random sample of $R$ 
realizations of the random vector $\vec{W}$. The SAA function is defined as
\begin{align}
  \hat{g}_R(\vec{x}) \triangleq \frac{1}{R} \sum_{r=1}^R G(\vec{x}, \vec{W}^r),
  \tag{11}
  \label{SAA_goal}
\end{align}
and the SAA problem $\mathcal{P}_3$ is defined as 
\begin{align}
  \mathcal{P}_3: & \min_{\vec{x} \in \mathcal{X}} \hat{g}_R(\vec{x}). \notag
\end{align}
By Monte Carlo Sampling, with support from \textit{the Law of Large Numbers} \cite{LawofLargeNumbers}, when $R$ is large 
enough, the optimal value of $\hat{g}_R(\vec{x})$ can converge to the optimal value of $g(\vec{x})$ with probability 
one (w.p.1). As a result, we only need to care about how to solve $\mathcal{P}_3$ as optimal as possible. 

\begin{algorithm}[h]
  \caption{SAA-based Redundant Placement (SAA-RP)}
  \begin{algorithmic}[1]
    \STATE {Choose initial sample size $R$ and $R'$ ($R' \gg R$)}
		\STATE {Choose the number of replications $L$ (indexed by $\mathcal{L}$)}
    \STATE {Set up a gap tolerance $\epsilon$}
    \FOR {$l$ = $1$ to $L$ \textbf{in parallel}}
      \STATE {Generate $R$ independent samples $\vec{W}^1_l, ..., \vec{W}^R_l$}
      \STATE {Call GASS to obtain the minimum value of $\hat{g}_R(\vec{x}_l)$ with the form of
        $\frac{1}{R} \sum_{r=1}^R G(\vec{x}_l, \vec{W}_l^r)$}
      \STATE {Record the optimal goal $\hat{g}_R(\hat{\vec{x}}^*_l)$ and the corresponding 
        variable $\hat{\vec{x}}^*_l$ returned from GASS}
    \ENDFOR
    \STATE {$\bar{v}_{R}^* \leftarrow \frac{1}{L} \sum_{l=1}^{L} \hat{g}_R(\hat{\vec{x}}^*_l)$}
    \FOR {$l$ = $1$ to $L$ \textbf{in parallel}}
      \STATE {Generate $R'$ independent samples $\vec{W}^1_l, ..., \vec{W}^{R'}_l$}
      \STATE {$v^l_{R'} \leftarrow \frac{1}{R'} \sum_{r'=1}^{R'} G(\hat{\vec{x}}^*_l, \vec{W}_l^{r'})$}
    \ENDFOR
    \STATE {Get the worst replication $v^{\bullet}_{R'} \leftarrow \max_{l \in \mathcal{L}} v^l_{R'}$}
    \IF {the gap $v^{\bullet}_{R'} - \bar{v}_{R}^* < \epsilon$}
    \STATE {Choose the best solution $\hat{\vec{x}}^*_l$ among all $L$ replicationss}
    \ELSE 
		\STATE {Increase $R$ (for drill) and $R'$ (for evaluation)}
		\STATE {\textbf{goto} Step. 4}
    \ENDIF
		\RETURN {the best solution $\hat{\vec{x}}^*_l$}
  \end{algorithmic}
  \label{SAA-RP}
\end{algorithm}

The SAA-RP framework is presented in \textbf{Algorithm \ref{SAA-RP}}.
Firstly, we need to select the sample size $R$ and $R'$ appropriately. As the sample size $R$ increases, the optimal solution of the 
SAA problem $\mathcal{P}_2$ converges to its \textit{true problem} $\mathcal{P}_1$. However, the computational complexity for solving 
$\mathcal{P}_2$ increases at least linearly, even exponentially, in the sample size $R$ \cite{SAA}. Therefore, when we 
choose $R$, the trade-off between the quality of the optimality and the computation effort should be taken into account. 
Besides, $R'$ here is used to obtain an estimate of $\mathcal{P}_1$ with the obtained solution of $\mathcal{P}_2$. 
In order to obtain an accurate estimate, we have every reason to choose a relatively large sample size $R'$ ($R' \gg R$). 
Secondlly, inspired by \cite{SAA}, we replicate generating and solving $\mathcal{P}_2$ with $L$ i.i.d. replications. From Step. 4 
to Step. 8, we call the algorithm GASS to obtain the asymptotically optimal solution of $\mathcal{P}_2$ and record the best-so-far results. From 
Step. 10 to Step. 13, we estimate the true value of $\mathcal{P}_1$ for each replication. After that, those estimates 
are compared with the average of those optimal solutions of $\mathcal{P}_2$. If the maximum gap is smaller than the tolerance, 
SAA-RP returns the best solution among $L$ replications and the algorithm terminates, otherwise we increase $R$ and $R'$ and drill 
again. 

\subsection{The GASS Algorithm}
GASS is implemented based on the well-konwn Genetic Algorithm (GA). The detailed procedure is demonstrated in \textbf{Algorithm \ref{GASS}}. 
Firstly, we initialize the necessary parameters, including 
the population size $P$, the number of iterations \texttt{it}, and the probability of crossover $\mathbb{P}_c$ and mutation $\mathbb{P}_m$, 
respectively. After that, we randomly generate the initial population from the domain $\mathcal{X}$. From Step. 6 to Step. 10, GASS 
executes the crossover operation. At the beginning of this operation, GASS checks whether crossover need to be executed. If yes, 
GASS choose the \textit{best} two chromosomes according to their fitness values. With that, the latter part of $\vec{x}_{p_1}$ and 
$\vec{x}_{p_2}$ are exchanged since the position $\vec{x}(b_j)$. From Step. 11 to Step. 13, GASS executes the mutation operation. At 
the beginning of this operation, it checks whether each chromosome can mutate according to the mutation probability $\mathbb{P}_m$. 
At the end, only the chromosome with the best fitness value can be returned. 

\begin{algorithm}
  \caption{GA-based Server Selection (GASS)}
  \begin{algorithmic}[1]
    \STATE {Initialize the population size $P$, number of iterations \texttt{it}, the probability of crossover 
    $\mathbb{P}_c$ and mutation $\mathbb{P}_m$}
    \STATE {Randomly generate $P$ chromosomes $\vec{x}_1, ..., \vec{x}_P \in \mathcal{X}$}
    \FOR {$t = 1$ to \texttt{it}}
      \STATE {$\forall p \in \{1, ..., P\}$, renew the optimization goal of $\mathcal{P}_2$, i.e. 
      $\hat{g}_R(\vec{x}_p)$, according to \eqref{SAA_goal}}
      \FOR {$p = 1$ to $P$}
        \IF {$\texttt{rand}() < \mathbb{P}_c$}
        \STATE {Choose two chromosomes $p_1$ and $p_2$ according to the probability distribution: 
        \begin{align*}
          \mathbb{P}(p \textrm{ is chosen}) = \frac { 1 / \hat{g}_R(\vec{x_p})} {\sum_{p'=1}^P 1 / \hat{g}_R(\vec{x_{p'}})}
        \end{align*}
        }
        \STATE {Randomly choose SBS $j \in \mathcal{M}$} 
        \STATE {Crossover the segement of $\vec{x}_{p_1}$ and $\vec{x}_{p_2}$ after the partitioning point $\vec{x}(b_{j-1})$: 
        \begin{align*}
          [\vec{x}_{p_1}(b_j), ..., \vec{x}_{p_1}(b_M)] \leftrightarrow [\vec{x}_{p_2}(b_j), ..., \vec{x}_{p_2}(b_M)]
        \end{align*}
        }
      \ENDIF
        \IF {$\texttt{rand}() < \mathbb{P}_m$}
          \STATE {Randomly choose SBS $j \in \mathcal{M}$ and re-generate the segement $\vec{x}_p(b_j)$}
        \ENDIF
      \ENDFOR
    \ENDFOR
    \RETURN {$\argmin_{p} \hat{g}(\vec{x_p})$ from $P$ chromosomes}
  \end{algorithmic}
  \label{GASS}
\end{algorithm}

\subsection{Strength and Advantages}\label{sec5.4}
This subsection summarizes the strength and advantages of SAA-RP and GASS. 

1) Notice that SAA-RP is not deployed online. It can be periodically re-run to follow up end users' service 
demand pattern. During each period, for example, a month or a quarter, end users' microservice composition 
preferences can be collected (under authorization of privacy). Pods can be reconstructed based on the result from 
SAA-RP. This process can be carried out through rolling upgrade with Kubernetes. 

2) With the recoded decision variable $\vec{x}$, GASS is simple to operate and it enjoys a fast convergence rate. 
In the domain of $\mathcal{P}_1$, the number of elements is $\exp \{ \sum_{q \in \mathcal{Q}} C_q \cdot \ln M \}$, 
which \textit{exponentially} increases with the scale of microservices. However, after re-encoding, the number of elements 
in domain $\mathcal{X}$ is $\prod_{j \in \mathcal{M}} \big(b_j (\sum_{q \in \mathcal{Q}} C_q + 1) - \frac{b_j}{2}(b_j-1) \big)$, 
which increases \textit{polynomially} with the scale of microservices. The conclusion will be proved in Subsection 
\ref{sec6.1} and verified in Subsection \ref{sec7.3.2}. 

3) SAA-RP makes up with the shortcoming of the default scheduler of Kubernetes when encountering the MEC.. \textit{Kube-scheduler} 
is a component responsible for the deployment of configured pods and microservices, which selects the node for a 
microservice instance in a two-step operation: Filtering and Scoring \cite{Kube-scheduler}. The filtering step 
finds the set of nodes who are feasible to schedule the microservice instance based on their available resources. 
In the scoring step, the \texttt{kube-scheduler} ranks the schedulable nodes to choose the most suitable one for 
the placement of the microservice instance. It places microservices only based on resources occupancy of nodes. 
By contrast, SAA-RP takes both the service request pattern of end users and the heterogeneity of the distributed 
nodes into consideration. SAA-RP makes a progress towards the resource orchestration on the heterogenous edge. 

\section{Theoretical Analysis}\label{sec6}
In this section, we analyze the optimality of SAA-RP and the complexity of GASS.

\subsection{Solution Optimality}\label{sec6.2}
Recall that the domain $\mathcal{X}$ of problem $\mathcal{P}_2$ and $\mathcal{P}_3$ is finite, whose size is 
$\prod_{j \in \mathcal{M}} b_j \big(\sum_{q \in \mathcal{Q}} C_q + 1 - \frac{1}{2}(b_j-1) \big)$. Thus, 
$\mathcal{P}_2$ and $\mathcal{P}_3$ have nonempty set of optimal solutions, denoted as $\mathcal{X}^*$ and 
$\hat{\mathcal{X}}_R$, respectively. We let $v^*$ and $\hat{v}_R$ denote the optimal values of $\mathcal{P}_2$ 
and $\mathcal{P}_3$, respectively. To analysis the optimality, we also define the set of $\epsilon$-optimal solutions. 
\begin{definition}
  \textbf{$\epsilon$-optimal Solutions}
  For $\epsilon \geq 0$, if $\vec{x} \in \mathcal{X}$ and $g(\vec{x}) \leq v^* + \epsilon$, then we say that $\vec{x}$ 
  is an $\epsilon$-optimal solution of $\mathcal{P}_1$. Similarly, if $\vec{x} \in \mathcal{X}$ and 
  $g(\vec{x}) \leq \hat{v}_R + \epsilon$, $\vec{x}$ is an $\epsilon$-optimal solution of $\mathcal{P}_3$.
\end{definition}
We use $\mathcal{X}^{\epsilon}$ and $\hat{\mathcal{X}}_R^{\epsilon}$ to denote the set of $\epsilon$-optimal solutions 
of $\mathcal{P}_2$ and $\mathcal{P}_3$, respectively. Then we have the following proposition. 
\begin{proposition}
  $\hat{v}_R \to v^*$ w.p.1 as $R \to \infty$; $\forall \epsilon \geq 0$, the event 
  $\{ \hat{\mathcal{X}}_R^{\epsilon} \to \mathcal{X}^{\epsilon} \}$ happens w.p.1 for $R$ large enough.
  \label{conv_goal}
\end{proposition}
\begin{proof}
  It can be directly obtained from \textit{Proposition 2.1} of \cite{SAA}, not tired in words here.
\end{proof}
Proposition \ref{conv_goal} implies that for almost every realization $\vec{\omega} = \{ \vec{W}^1, \vec{W}^2, ... \}$ 
of the random vector, there exists an integer $R(\vec{\omega})$ such that $\hat{v}_R \to v^*$ and 
$\{ \hat{\mathcal{X}}_R^{\epsilon} \to \mathcal{X}^{\epsilon} \}$ happen for all samples $\{\vec{W}^1, ..., \vec{W}^r\}$ 
from $\vec{\omega}$ with $r \geq R(\vec{\omega})$. 

The following proposition demonstrates the convergence rate of SAA method.
\begin{proposition}
  $\forall \epsilon > 0$ small enough and $\varpi \in [0, \epsilon)$, for the probability 
  $\mathbb{P}(\{ \hat{\mathcal{X}}_R^{\varpi} \subset \mathcal{X}^{\epsilon} \})$ to be at least $1 - \alpha$, the number of 
  sample size $R$ should satisify 
  \begin{align*}
    R \geq \frac{3 \sigma^2_{max}}{(\epsilon - \varpi)^2} \cdot 
    \sum_{j \in \mathcal{M}} \log \Big( \frac{b_j}{\alpha} \cdot \big(\sum_{q \in \mathcal{Q}} C_q + \frac{3 - b_j}{2} \big) \Big),  
  \end{align*}
  where 
  \begin{align*}
    \sigma^2_{max} \triangleq \max_{\vec{x} \in \mathcal{X} \backslash \mathcal{X}^{\epsilon}} {\rm{Var}} \big[G(u(\vec{x}), \vec{W}) - G(\vec{x}, \vec{W})\big],
  \end{align*} 
  and $u(\cdot)$ is a mapping from $\mathcal{X} \backslash \mathcal{X}^{\epsilon}$ into $\mathcal{X}^*$, which satisifies that 
  $\forall \vec{x} \in \mathcal{X} \backslash \mathcal{X}^{\epsilon}, g(u(\vec{x})) \leq g(\vec{x}) - v^*$. 
  \label{rate}
\end{proposition}
\begin{proof}
  It can be directly obtained by combing \textit{Proposition 2.2} of \cite{SAA} with the size of $\mathcal{X}$. For 
  more details, please consult \cite{SAA} directly.
\end{proof}
Proposition \ref{rate} implies that the number of samples $R$ depends \textit{logarithmically} on the feature of domain 
$\mathcal{X}$ and the tolerance probability $\alpha$. 

\subsection{Algorithm Complexity}\label{sec6.1}
Obviously, GASS works through a mechanism of decomposition and reassembly. The following proposition holds. 
\begin{proposition}
  For problem $\mathcal{P}_2$ and $\mathcal{P}_3$, when the number of SBSs is fixed as $M$, no matter how many mobile devices 
  there are, the convergence time of GASS is 
  $O \Big(\sqrt{ R \cdot \prod_{j \in \mathcal{M}} \big(b_j (\sum_{q \in \mathcal{Q}} C_q + 1) - \frac{b_j}{2}(b_j-1) \big)} \Big)$. 
  \label{complexity}
\end{proposition}
\begin{proof}
  Notice that each candidate only has to be deployed once on a single SBS. Thus, the number of possible 
  values of $\vec{x}(b_j)$ is $b_j (\sum_{q \in \mathcal{Q}} C_q + 1) - \frac{b_j}{2}(b_j-1)$, and the number 
  of elements in domain $\mathcal{X}$, i.e. the problem size, is 
  $R \cdot \prod_{j \in \mathcal{M}} \big(b_j (\sum_{q \in \mathcal{Q}} C_q + 1) - \frac{b_j}{2}(b_j-1) \big)$. 
  It follows from the fact that the $k$th element of $\vec{x}(b_j)$ has $\sum_{q \in \mathcal{Q}} C_q + 2 -k$ choices. 
  Considering the processing building blocks of $\mathcal{P}_3$ is of equal salience, the result can be obtained 
  directly \cite{searching-book} \cite{GA-complexity}.
\end{proof}
Proposition \ref{complexity} indicates that the complexity of GASS increases polynomially with the scale of 
the application, i.e., $\sum_{q \in \mathcal{Q}} C_q$. 

\section{Experimental Evaluation}\label{sec7}
In this section, we verify the superiority of the proposed algorithms through simulations. 

\subsection{Benchmark Policies}\label{sec7.1}
Our method is compared with several representative baselines and a state-of-the-art algorithm, GenDoc \cite{placement-add1}. 
The baselines are performed in two scenarios while GenDoc is performed as it was defined in \cite{placement-add1}. In the first scenario, 
redundancy is not allowed. Each candidate can only be dispatched to only one SBS. It is used to evaluate the superiority of redundant 
placement. In the second scenario, redundancy is allowed. It is used to evaluate the optimality of GASS. Those benchmark policies, including 
GenDoc, are used to replace GASS to generate the best-so-far solution of each sampling. In both scenarios, those benchmark policies will be run $R$ 
times and the average value is returned. Details are summarized as follows.

(1) \textbf{Random Placement in Scenario \#1 (RP1)}: $\forall q \in \mathcal{Q}, \forall c_q \in \mathcal{C}_q$, 
dispatch $c_q$ to a randomly chosen SBS $j$ and decrement $b_j$. The procedure terminates if no available SBS.

(2) \textbf{Random Placement in Scenario \#2 (RP2)}: $\forall q \in \mathcal{Q}, \forall c_q \in \mathcal{C}_q$, 
randomly dispatch $c_q$ to $m$ SBSs. The number $m \in \{m' \in \mathbb{N} | 0 \leq m' \leq M\}$ is generated randomly. After 
that, for those SBSs, decrement their $b_j$. The procedure terminates if no available SBS. 

(3) \textbf{Genetic Algorithm in Scenario \#1 (GA1)}: The chromosome is encoded as $[p(s_1^1), ..., p(s_Q^{C_Q})]^\top$, where $p(s)$ 
is the chosen SBS for the placement of the candidate $s$. This encoding ensures that each candidate can only be dispatched to one SBS. 
Based on that, each generation of chromosomes are created by selection, recombination, and mutation. 

(4) \textbf{Greedy Placement in Scenario \#2 (GP2)}: For each SBS $j \in \mathcal{M}$, $b_j$ candidates will be deployed on it. 
It means that the feasible region lies in the boundary of the constraint \eqref{cons_1}. In each iteration, each end user always chooses the 
nearest available SBS to execute its selected candidates. 

(5) \textbf{GenDoc}: GenDoc is a configuration-aware placement and scheduling algorithm, proposed in \cite{placement-add1}. To apply GenDoc 
to our system model, $\forall j \in \mathcal{M}$, we set $C_j^{vir} = b_j$, where $C_j^{vir}$ is the maximum virtual capacity 
of the $j$th SBS. More details can be found in Section 4.2 of \cite{placement-add1}.

\subsection{Experimental Settings}\label{sec7.2}
All the experiments are implemented in MATLAB R2019b on macOS Catalina equipped with 3.1 GHz Quad-Core Intel Core i7 and 16 GB RAM. The 
parameter settings are discussed as follows. 

\begin{table}[!ht]
  \renewcommand{\arraystretch}{1.2}
  \caption{Parameter settings.}
  \label{table2}
  \centering
  \begin{tabular}{c c | c c}
    \hline\hline
    \textbf{Parameter} & \textbf{Value} & \textbf{Parameter} & \textbf{Value}\\
    \hline
    $Q$ & $10$ & $\forall q, C_q$ & $[2, 5]$ \\[+0.5mm]
    $N$ & $500$ & $M$ & $40$ \\[+0.5mm]
    $b_l$ & $3$ & $b_u$ & $5$ \\[+0.5mm]
    $\alpha$ & $[1,8]$ kbits & $\beta$ & $5$ ms \\[+0.5mm]
    $\tau_b$ & $0.1$ s & $L$ & $5$ \\[+0.5mm]
    $\forall s, \tau_{exe}(j_p(s))$ & $[1, 2]$ ms & signal radius & $[200, 600]$ m \\[+0.5mm]
    \hline
    $R$ & $500$ & $R'$ & $100000$ \\[+0.5mm]
    $L$ & $10$ & $\epsilon$ & $2 \times 10^{-4}$ \\[+0.5mm]
    $P$ & $10$ & \texttt{it} & $300$ \\[+0.5mm]
    $\mathbb{P}_m$ & $10\%$ & $\mathbb{P}_c$ & $80\%$ \\[+0.5mm]
    \hline\hline
  \end{tabular}
\end{table}
\textit{The microservices and candidates:} The number of microservices in the application $Q$ is set as $10$ in default. 
For each microservice $q$, the number of its candidates is uniformly sampled from the integer interval $[2,5]$. In each replication 
$\vec{W}^r$, the service composition scheme of the $i$th end user is sampled according to $\mathbb{P}(E(\vec{s}(i)))$. Considering that 
there is no commonly used dataset for microservice composition, $\forall q \in \mathcal{Q}, \forall c \in \mathcal{C}_q$, we generate 
$\mathbb{P}(E(s_q^c))$ uniformly to avoid any bias. 

\textit{The pre-5G HetNet:} The experiment is conducted based on the geolocation information of base stations and end users within the Melbourne 
CBD area contained in the EUA dataset \cite{EUA}. In our simulations, we choose 500 end users and 40 SBSs uniformly from the
dataset in default. The coverage radius of each SBS is sampled from $[200,600]$ meters uniformly. 
$\forall i \in \mathcal{N}$, $\forall j \in \mathcal{M}$, $\frac{1}{d(i, j)} = 1$ MHz.
In addition, the maximum hops 
between any two SBSs can not larger than $4$. $\forall j \in \mathcal{M}$, $b_j$ is chosen from the integer interval $[b_l, b_u]$. 
We set $b_l = 3, b_u = 5$ in default. 

\begin{figure*} 
  \centering 
  \subfigure[The convergence rate of all the algorithms with $N = 500$, $M = 40$.]{ 
    \label{fig4a} 
    \includegraphics[height=2.3in]{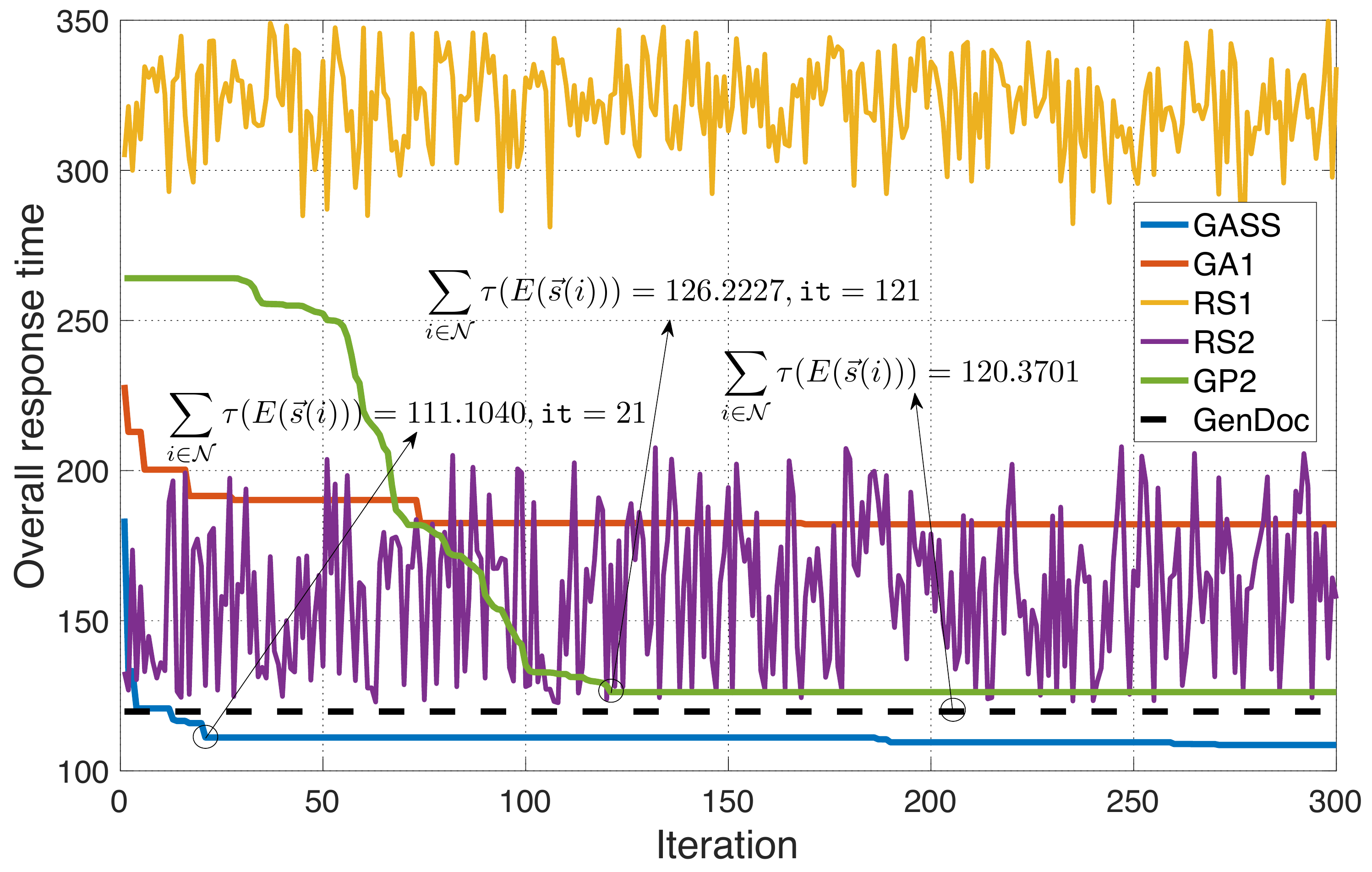}
  }
  \subfigure[The overall response time vs. $\overline{b}$.]{ 
      \label{fig4b} 
      \includegraphics[height=2.3in]{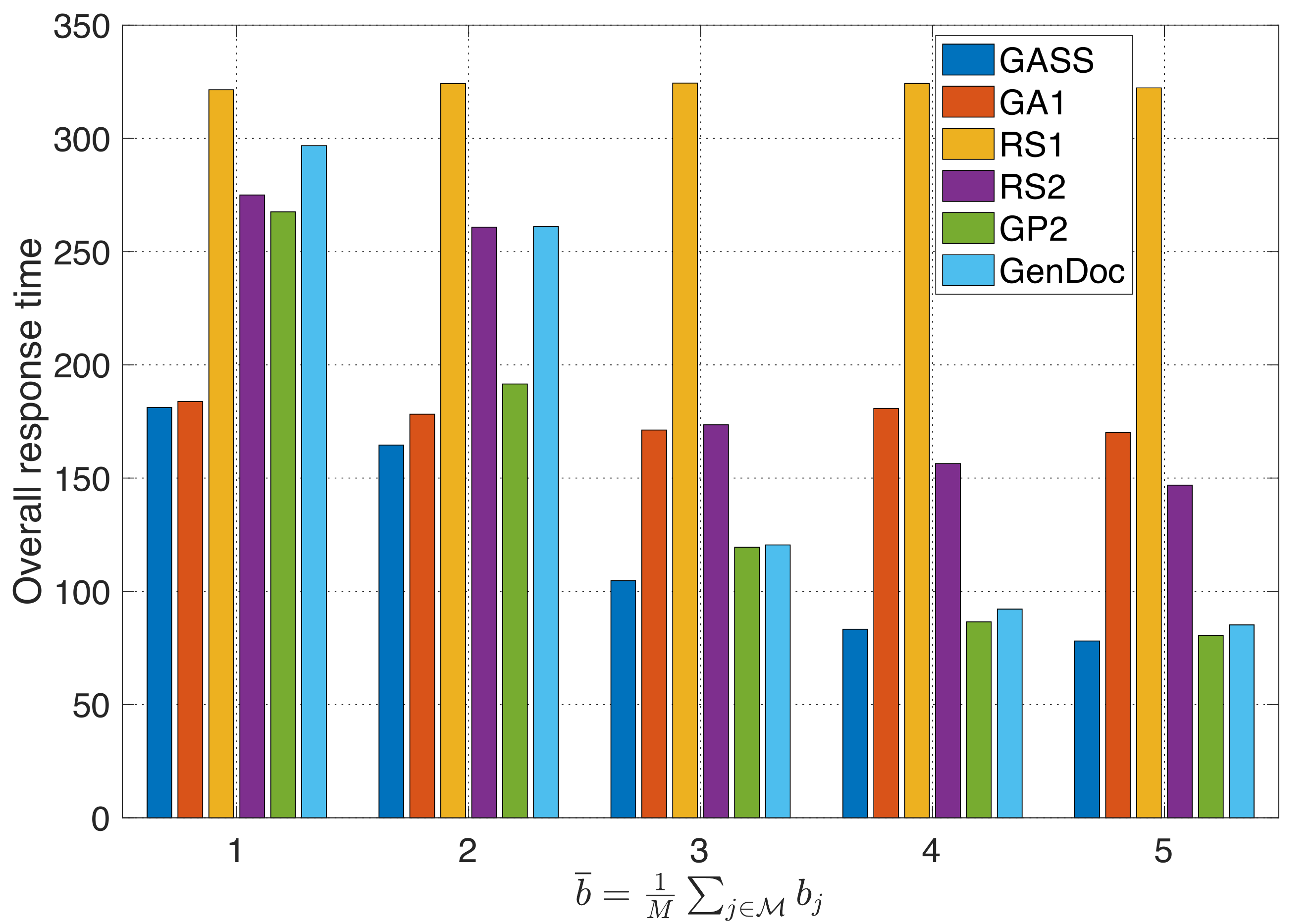}
    } 
  \caption{Algorithm performace comparison.} 
  \label{fig:subfig} 
\end{figure*}

\subsection{Experiment Results}\label{sec7.3}
The experiments are conducted to analyze the optimality and scalability of the proposed algorithm. 

\subsubsection{Optimality}
As shown in Fig. \ref{fig4a}, GASS outperforms all the other algorithms in the overall response time, i.e., 
$\sum_{i \in \mathcal{N}} \tau(E (\vec{s}(i)))$. Specifically, within $300$ iterations, GASS outperforms 
GenDoc, GP2, RS2, GA1, and RS1 by $10.81\%$, $16.20\%$, $44.81\%$, $67.66\%$, and $196.43\%$, respectively. 
The result verifies both the rationality of redundant placement and the optimality of our algorithm. As for 
the former, it is verified by that all the algorithms implemented in scenario \#2 perform better than the 
algorithms implemented in scenario \#1. The superiority of redundant placement lies in that it takes full 
advantage of the distributed SBSs' limited resources. In this case, the processing of end users' 
service requests can surely be balanced. As for the latter, it is verified by that GASS can converge to an 
approximate optimal solution, i.e. $111.1040$ ms at a very rapid rate. The solution achieved at the $21^{th}$ 
iteration is already better than all the other algorithms. 

In addition to the above phenomena, it is interesting to see that GP2 can obtain a relatively good result. 
We can verify that for each deployment, what GP2 adopts is the optimal operation. For each microservice, 
only the most frequently requested candidate has the privilege to be deployed, which ensures that the maximum 
number of mobile devices can enjoy their optimal situation. In comparison, GenDoc consists of \textit{greedy} 
placement (server configuration) and \textit{dynamic programming}-based microservice scheduling, which is not 
overbearing. 

Fig. \ref{fig4b} shows that GASS can keep on top under different conditions. In this figure, the horizontal axis 
is the mean value of can-be-deployed candidates of each SBS $j$, i.e. $b_j$. We can find that for those algorithms 
which are implemented in scenario \#1, $\overline{b}$ has no significant effect on their performance. The reason 
is that whatever $b_j$ takes, only one candidate can be deployed on each SBS $j$. It is also why GA1 can achieve 
a similar result with GASS when $\overline{b}=1$. By contrast, with the increase in $\overline{b}$, all the 
algorithms implemented in scenario \#2 enjoy less response time. The result is obvious because $b_j$ decides 
the upper limit of resources, which is the key influence factor of time consumption. It is also worth noting 
that when $\overline{b}$ increases, the gap between GASS, GP2, and GenDoc is likely to narrow. This is exactly 
the embodiment of the trade-off between algorithm improvement and resource promotion. When resources are rich, 
even poorly performing algorithms can produce good results. 

\subsubsection{Scalability}\label{sec7.3.2}
Scalability is embodied in two aspects, the HetNet and the application (microservices and candidates). 

\begin{figure}[!t]
  \centering
  \includegraphics[width=3.3in]{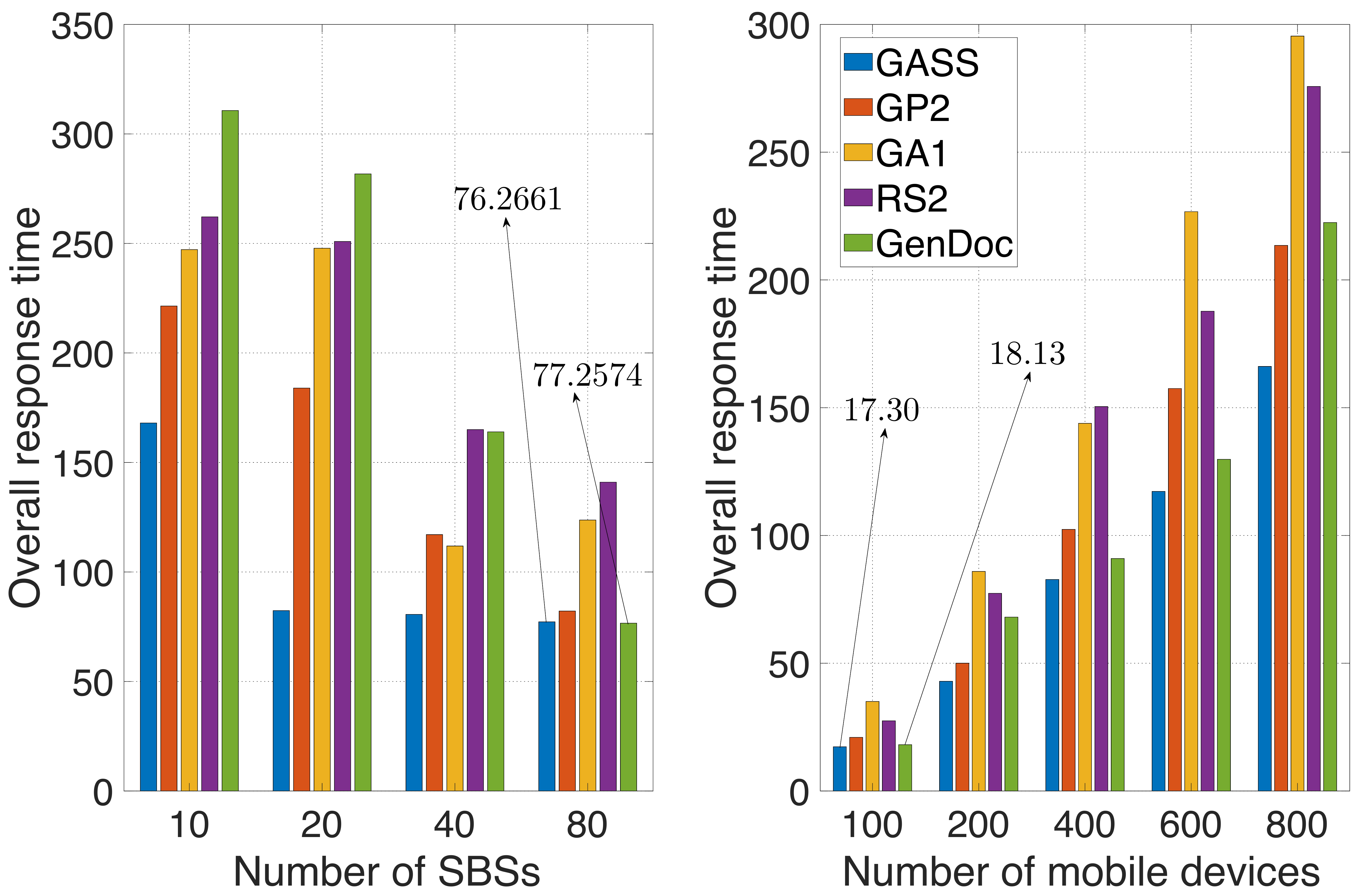}
  \caption{Overall response time vs. scale of the HetNet.}
  \label{fig5}
\end{figure}

\textbf{\textit{The HetNet:}} the scale of the HetNet is embodied in two variables, the number of SBSs $M$ and 
the number of mobile devices $N$. The left of Fig. \ref{fig5} demonstrates the impact of $M$ on the overall response 
time. Generally, as $M$ increases, the response time decreases. This is because more SBSs can provide more 
resources, which greatly helps to realize \textit{near-request processing}. Even so, the superiority of GASS is 
obvious, as it is always the best of five algorithms whatever $M$ takes (RS1 is discarded). Thus, the scalability 
of GASS holds. Besides, there are some noteworthy phenomena. Firstly, the response time of GA1 has a slightly 
rising trend when $M$ increases from $40$ to $80$. This is because when $M$ increases, the dimension of feasible 
solution increases, which greatly expands the solution space. Meanwhile, the connected graph of SBSs become sparse, 
which leads to more hops to transfer data streams. Under this circumstances, $300$ iterations might not be enough 
to achieve an optimal enough solution. However, GASS is not effected because the solution space of GASS is much 
smaller. The phenomenon verifies the second advantage displayed in Subsection \ref{sec5.4}. Secondly, when $M$ 
increases, the gap between GASS, GP2, and GenDoc is likely to narrow. This phenomenon has been captured in Fig. 
\ref{fig4b}. No matter increasing $M$ or $\overline{b}$, the resources of SBSs are increasing, and poorly 
performing algorithms can produce good results. 

The right of Fig. \ref{fig5} demonstrates the impact of $M$ on the overall response time. For all the implemented 
algorithms, the overall response time increases as $N$ increases while the solution achieved by GASS is always the 
best. It is interesting that the gaps between those benchmark policies and GASS increase as $N$ increases. It indicates that 
GASS is robust to the computation complexity of the fitness function. Thus, the scalability of GASS holds. 

\begin{figure}[!t]
  \centering
  \includegraphics[width=2.7in]{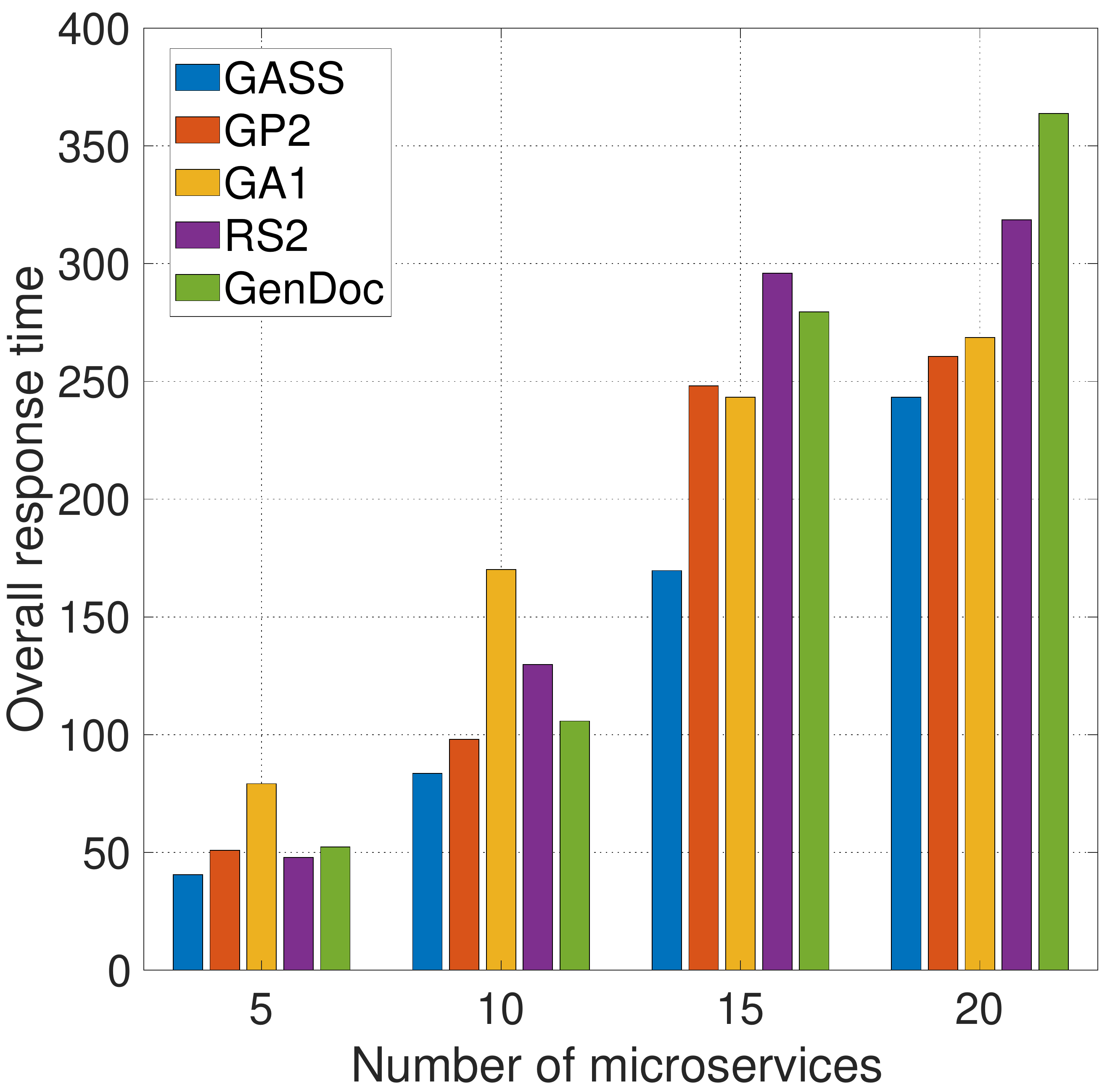}
  \caption{Overall response time vs. number of the microservices.}
  \label{fig6}
\end{figure}

\begin{figure}[!t]
  \centering
  \includegraphics[width=3.5in]{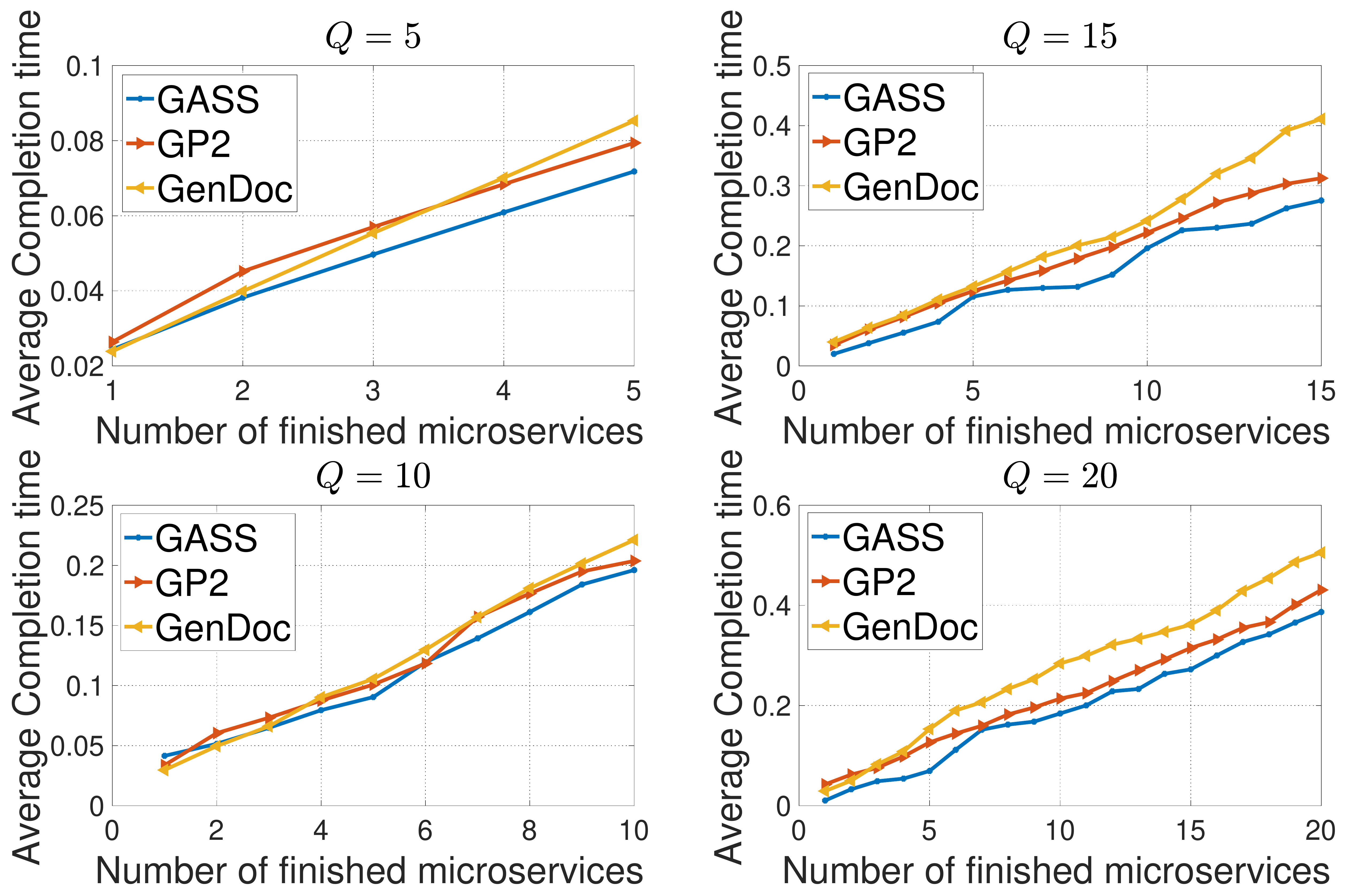}
  \caption{Average response time vs. number of the microservices.}
  \label{fig7}
\end{figure}

\textbf{\textit{The application:}} the scale of the application is embodied in two variables, the number of microservices 
$Q$ and the average number of candidates per microservice $\overline{C} \triangleq \frac{1}{Q} \sum_{q \in \mathcal{Q}} C_q$. 
It can be concluded that GenDoc and GP2 are competitive while RS1, RS2 and GA1 are obviously lagging behind. 
Thus, in the following analysis, we only compare GASS with GenDoc and GP2 in terms of the average completion 
time. 

\begin{figure}[!t]
  \centering
  \includegraphics[width=2.5in]{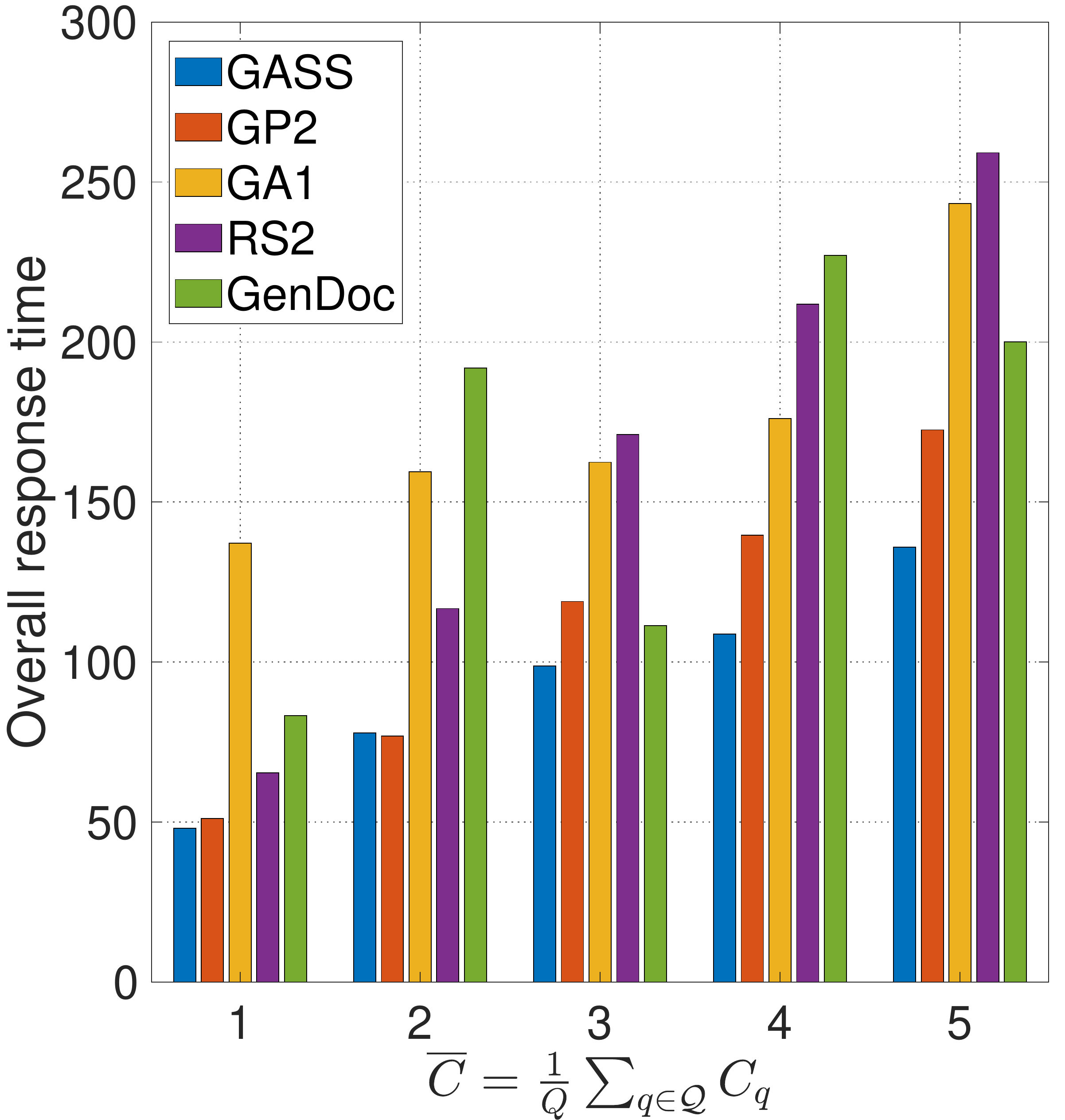}
  \caption{Overall response time vs. $\overline{C}$.}
  \label{fig8}
\end{figure}

\begin{figure}[!t]
  \centering
  \includegraphics[width=3.5in]{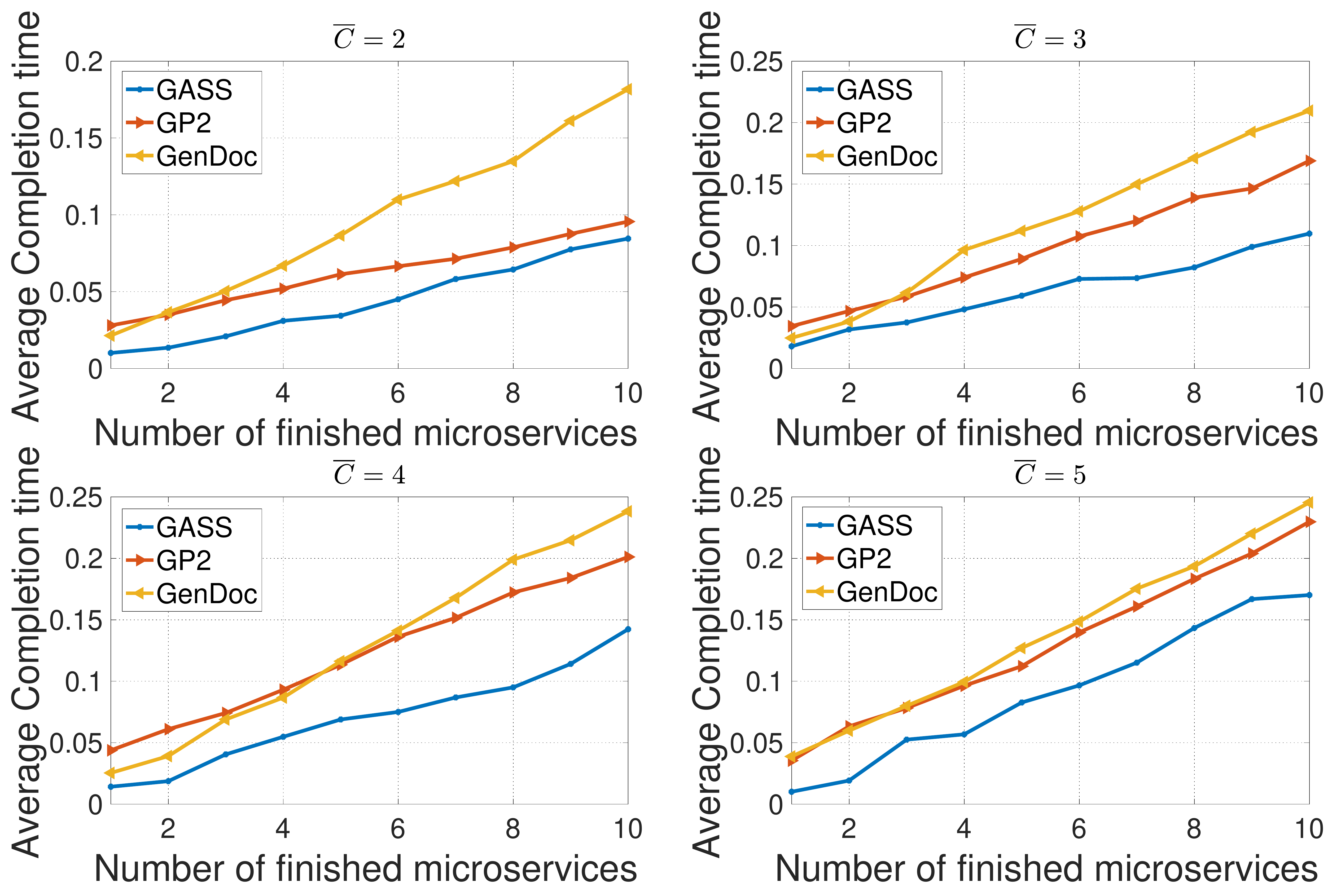}
  \caption{Average response time vs. $\overline{C}$.}
  \label{fig9}
\end{figure}

Fig. \ref{fig6} and Fig. \ref{fig7} demonstrates the impact of the scale of microservices. From Fig. \ref{fig6} 
we can find that GASS can keep on top whatever $Q$ is. Correspondingly, 
Fig. \ref{fig7} demonstrates the involution of the average completion time per mobile device when $Q$ is $5$, 
$10$, $15$, and $20$, respectively. In all cases, GASS achieves the minimum average completion time no matter 
how many microservices have been finished. In our experiments, the maximum $\mathbb{E}[\sum_{q \in \mathcal{Q}} C_q]$ 
is $70$ while $\mathbb{E} [\sum_{j \in \mathcal{M}} b_j]$ is 160. Theoretically, if the expected number of 
all candidates $\mathbb{E}[\sum_{q \in \mathcal{Q}} C_q]$ does not exceed the expected number of all the 
can-be-deployed candidates $\mathbb{E} [\sum_{j \in \mathcal{M}} b_j]$, GASS can maintain its competitive edge. 
This is because no requests from end users need to be processed by cloud, and the time-consuming backbone can be saved. This 
advantage is not hold by the benchmark policies.
Fig. \ref{fig8} and Fig. \ref{fig9} demonstrates the impact of the scale of candidates. Similarly, from Fig. \ref{fig8}, 
we can find that GASS outperforms GP2 and GenDoc in most cases. From Fig. \ref{fig9} we can find that 
GASS always achieves the minimum average completion time when $\overline{C}$ is $2$, $3$, $4$, and $5$. 

Fig. \ref{fig6} $\sim$ Fig. \ref{fig9} verifies that the scalability of GASS holds in terms of the number of microservices and 
candidates. 

\begin{table}[!ht]
  \renewcommand{\arraystretch}{1.2}
  \caption{Impact of population size, crossover probability, and mutation probability.}
  \label{table3}
  \centering
  \begin{tabular}{c c | c c | c c}
    \hline\hline
    $P$ & $\sum_{i \in \mathcal{N}} \tau_i$ & $\mathbb{P}_c$ & $\sum_{i \in \mathcal{N}} \tau_i$ & $\mathbb{P}_m$ & $\sum_{i \in \mathcal{N}} \tau_i$\\[+0.5mm]
    \hline
    $5$  & $114.6762$ & $20\%$ & $114.5866$ & $20\%$ & $110.4808$ \\[+0.5mm]
    $10$ & $109.9742$ & $40\%$ & $112.2296$ & $40\%$ & $105.8360$ \\[+0.5mm]
    $15$ & $109.6814$ & $60\%$ & $116.8252$ & $60\%$ & $112.7771$ \\[+0.5mm]
    $20$ & $113.4527$ & $80\%$ & $111.4439$ & $80\%$ & $109.2450$ \\[+0.5mm]
    $25$ & $111.8812$ & $100\%$ & $116.1299$ & $100\%$ & $112.6478$ \\[+0.5mm]
    \hline\hline
  \end{tabular}
\end{table}

\textbf{\textit{Ths superparameters of GASS:}}
Table \ref{table3} demonstrates the overall response time of GASS under different population size, probability 
of mutation $\mathbb{P}_m$, and probability of crossover $\mathbb{P}_c$, respectively. We can find 
that their impact is minor on the optimality of GASS. As a result, no more detailed discussion is launched. 

\section{Related Works}\label{sec2}
Service computing based on traditional cloud data-centers has been extensively studied in the last several 
years, especially service selection for composition \cite{composition3}\cite{review-add2}\cite{review-add4}, 
service provision\cite{provision1}, discovery \cite{discovery1}, and so on. 
However, putting \textit{everything about services} onto the distributed and heterogenous edge is still an 
area waiting for exploration. Multi-access Edge computing, as a increasingly popular computation paradigm, 
is facing the transition from theory to practice. The key to the transition is the placement and deployment 
of service instances. 

In the last two years, service placement at the distributed edge has been tentatively explored from the perspective 
of Quality of Experience (QoE) of end users or the budget of ASPs. For example, Ouyang et al. study the problem in 
a mobility-aware and dynamic way \cite{placement-follow-me}. Their goal is to optimize the long-term time averaged 
migration cost triggered by \textit{user mobility}. They develop two efficient heuristic schemes based on the Markov 
approximation and best response update techniques to approach a near-optimal solution. System stability is also 
guaranteed by Laypunov optimization technique. Chen et al. study the problem in a spatio-temporal way, under a limited 
budget of ASPs \cite{placement-MAB}. They pose the problem as a combinatorial contextual bandit learning problem and 
utilize Multi-armed Bandit (MAB) theory to \textit{learn} the optimal placement policy. However, the proposed algorithm 
is time-consuming and faces the curse of dimensionality. Except for the typical examples listed above, there also 
exist works dedicated on joint resource allocation and load balancing in service placement \cite{placement-2} 
\cite{placement-3} \cite{placement-1} \cite{review-add3}. However, as we have mentioned before, those works only study the to-be-placed 
services in an atomic way. The correlated and composite property of services is not taken into consideration. Besides, 
those works do not tell us how to apply their algorithms to the service deployment in a practical system. To address 
these deficiencies, we navigate the service placement and deployment from the view of production practices. 
Specifically, we adopt redundant placement to the correlated microservices, which can be unified managed by Kubernetes. 

The idea of redundancy has been studied in parallel-server systems and computing clusters \cite{redundancy--1} 
\cite{redundancy--2} \cite{redundancy--3}. The basic idea of redundancy is dispatching the same job to multiple 
servers. The job is considered done as soon as it completes service on any one server \cite{Redundancy-model}. 
Typical job redundancy model is the $S\&X$ model, where $X$ is the job's inherent size, and $S$ is the server 
slowdown. It is designed based on the weakness of the traditional Independent Runtimes (IR) model, where a job’s 
replicas experience independent runtimes across servers. Unfortunately, although the $S\&X$ model indeed captures the practical 
characteristics of real systems, it still face great challenges to put it into use in service deployment at the edge 
bacause the geographically distribution and heterogeneity of edge sites are not considered. To solve the problem, 
in this paper we redesign the entire model while the idea of redundancy is kept.

This work significantly extends our preliminary work \cite{similar}. To improve the practicability, we analyze the 
response time of each mobile device in a more rigorous manner, and improve it by always finding the nearest available 
edge site. We also take the uncertainty of end users' service composition scheme into consideration. It greatly 
increases the complexity but is of signality. Most important of all, we embedd the idea of redundancy into the 
problem and design an algorithm with a faster convergence rate. 

\section{Conclusion}\label{sec8}
In this paper, we study a redundant placement policy for the deployment of microservice-based applications 
at the distributed edge. We first demonstrate the typical HetNet in the near future, and then explore the 
possibilities of the deployment of composite microservices with containers and Kubernetes. After that, 
we model the redundant placement as a stochastic optimization problem. For the application with composite 
and correlated microservices, we design the SAA-based framework SAA-RP and the GASS algorithm to dispatch 
microservice instances into edge sites. By creating multiple access to services, our policy boosts a faster 
response for mobile devices significantly. SAA-RP not only take the uncertainty of microservice composition 
schemes of end users, but also the heterogeneity of edge sites into consideration. The experimental results 
based on a real-world dataset show both the optimality of redundant placement and the efficiency of GASS. 
In addition, we give guidance on the implementation of SAA-RP with Kubernetes. In our future work, we will 
hammer at the implementation of the redundant deployment of complex DAGs with arbitrary shape. 
\ifCLASSOPTIONcompsoc
  \section*{Acknowledgments}
\else
  \section*{Acknowledgment}
\fi
This research was partially supported by the National Key Research and Development Program of 
China (No. 2017YFB1400601), Key Research and Development Project of Zhejiang Province (No. 
2017C01015), National Science Foundation of China (No. 61772461), Natural Science Foundation 
of Zhejiang Province (No. LR18F020003 and No.LY17F020014).

\bibliographystyle{IEEEtran}
\bibliography{IEEEabrv,ref.bib}

\begin{thebibliography}{10}
\providecommand{\url}[1]{#1}
\csname url@samestyle\endcsname
\providecommand{\newblock}{\relax}
\providecommand{\bibinfo}[2]{#2}
\providecommand{\BIBentrySTDinterwordspacing}{\spaceskip=0pt\relax}
\providecommand{\BIBentryALTinterwordstretchfactor}{4}
\providecommand{\BIBentryALTinterwordspacing}{\spaceskip=\fontdimen2\font plus
\BIBentryALTinterwordstretchfactor\fontdimen3\font minus
  \fontdimen4\font\relax}
\providecommand{\BIBforeignlanguage}[2]{{%
\expandafter\ifx\csname l@#1\endcsname\relax
\typeout{** WARNING: IEEEtran.bst: No hyphenation pattern has been}%
\typeout{** loaded for the language `#1'. Using the pattern for}%
\typeout{** the default language instead.}%
\else
\language=\csname l@#1\endcsname
\fi
#2}}
\providecommand{\BIBdecl}{\relax}
\BIBdecl

\bibitem{MEC_survey1}
Y.~{Mao}, C.~{You}, J.~{Zhang}, K.~{Huang}, and K.~B. {Letaief}, ``A survey on
  mobile edge computing: The communication perspective,'' \emph{IEEE
  Communications Surveys Tutorials}, vol.~19, no.~4, pp. 2322--2358,
  Fourthquarter 2017.

\bibitem{MEC_survey2}
T.~{Taleb}, K.~{Samdanis}, B.~{Mada}, H.~{Flinck}, S.~{Dutta}, and
  D.~{Sabella}, ``On multi-access edge computing: A survey of the emerging 5g
  network edge cloud architecture and orchestration,'' \emph{IEEE
  Communications Surveys Tutorials}, vol.~19, no.~3, pp. 1657--1681,
  thirdquarter 2017.

\bibitem{review-add1}
S.~{Deng}, H.~{Zhao}, W.~{Fang}, J.~{Yin}, S.~{Dustdar}, and A.~Y. {Zomaya},
  ``Edge intelligence: The confluence of edge computing and artificial
  intelligence,'' \emph{IEEE Internet of Things Journal}, pp. 1--1, 2020.

\bibitem{Docker}
\BIBentryALTinterwordspacing
``Docker: Modernize your applications, accelerate innovation.'' [Online].
  Available: \url{https://www.docker.com/}
\BIBentrySTDinterwordspacing

\bibitem{Kubernetes}
\BIBentryALTinterwordspacing
``Kubernetes: Production-grade container orchestration.'' [Online]. Available:
  \url{https://kubernetes.io/}
\BIBentrySTDinterwordspacing

\bibitem{placement-follow-me}
T.~{Ouyang}, Z.~{Zhou}, and X.~{Chen}, ``Follow me at the edge: Mobility-aware
  dynamic service placement for mobile edge computing,'' \emph{IEEE Journal on
  Selected Areas in Communications}, vol.~36, no.~10, pp. 2333--2345, Oct 2018.

\bibitem{placement-2}
T.~{He}, H.~{Khamfroush}, S.~{Wang}, T.~{La Porta}, and S.~{Stein}, ``It's hard
  to share: Joint service placement and request scheduling in edge clouds with
  sharable and non-sharable resources,'' in \emph{2018 IEEE 38th International
  Conference on Distributed Computing Systems (ICDCS)}, July 2018, pp.
  365--375.

\bibitem{placement-3}
B.~{Gao}, Z.~{Zhou}, F.~{Liu}, and F.~{Xu}, ``Winning at the starting line:
  Joint network selection and service placement for mobile edge computing,'' in
  \emph{IEEE INFOCOM 2019 - IEEE Conference on Computer Communications}, April
  2019, pp. 1459--1467.

\bibitem{placement-MAB}
L.~{Chen}, J.~{Xu}, S.~{Ren}, and P.~{Zhou}, ``Spatio–temporal edge service
  placement: A bandit learning approach,'' \emph{IEEE Transactions on Wireless
  Communications}, vol.~17, no.~12, pp. 8388--8401, Dec 2018.

\bibitem{placement-1}
F.~{Ait Salaht}, F.~{Desprez}, A.~{Lebre}, C.~{Prud'homme}, and
  M.~{Abderrahim}, ``Service placement in fog computing using constraint
  programming,'' in \emph{2019 IEEE International Conference on Services
  Computing (SCC)}, July 2019, pp. 19--27.

\bibitem{placement-our-work}
Y.~{Chen}, S.~{Deng}, H.~{Zhao}, Q.~{He}, Y.~{Li}, and H.~{Gao},
  ``Data-intensive application deployment at edge: A deep reinforcement
  learning approach,'' in \emph{2019 IEEE International Conference on Web
  Services (ICWS)}, July 2019, pp. 355--359.

\bibitem{placement-add1}
\BIBentryALTinterwordspacing
L.~Liu, H.~Tan, S.~H.-C. Jiang, Z.~Han, X.-Y. Li, and H.~Huang, ``Dependent
  task placement and scheduling with function configuration in edge
  computing,'' in \emph{Proceedings of the International Symposium on Quality
  of Service}, ser. IWQoS ’19.\hskip 1em plus 0.5em minus 0.4em\relax New
  York, NY, USA: Association for Computing Machinery, 2019. [Online].
  Available: \url{https://doi.org/10.1145/3326285.3329055}
\BIBentrySTDinterwordspacing

\bibitem{Kubernetes-redundancy-1}
L.~A. {Vayghan}, M.~A. {Saied}, M.~{Toeroe}, and F.~{Khendek}, ``Deploying
  microservice based applications with kubernetes: Experiments and lessons
  learned,'' in \emph{2018 IEEE 11th International Conference on Cloud
  Computing (CLOUD)}, July 2018, pp. 970--973.

\bibitem{Kubernetes-redundancy-2}
L.~A. Vayghan, M.~A. Saied, M.~Toeroe, and F.~Khendek, ``Kubernetes as an
  availability manager for microservice applications,'' \emph{CoRR}, vol.
  abs/1901.04946, 2019.

\bibitem{Redundancy-model}
K.~{Gardner}, M.~{Harchol-Balter}, A.~{Scheller-Wolf}, and B.~{Van Houdt}, ``A
  better model for job redundancy: Decoupling server slowdown and job size,''
  \emph{IEEE/ACM Transactions on Networking}, vol.~25, no.~6, pp. 3353--3367,
  Dec 2017.

\bibitem{Kube-scheduler}
\BIBentryALTinterwordspacing
``Kubernetes scheduler.'' [Online]. Available:
  \url{https://kubernetes.io/docs/concepts/scheduling/kube-scheduler/}
\BIBentrySTDinterwordspacing

\bibitem{EUA}
P.~Lai, Q.~He, M.~Abdelrazek, F.~Chen, J.~Hosking, J.~Grundy, and Y.~Yang,
  ``Optimal edge user allocation in edge computing with variable sized vector
  bin packing,'' in \emph{Service-Oriented Computing}.\hskip 1em plus 0.5em
  minus 0.4em\relax Cham: Springer International Publishing, 2018, pp.
  230--245.

\bibitem{network_flows}
R.~K. Ahuja, T.~L. Magnanti, and J.~B. Orlin, ``Network flows,'' 1988.

\bibitem{network_slicing}
X.~{Foukas}, G.~{Patounas}, A.~{Elmokashfi}, and M.~K. {Marina}, ``Network
  slicing in 5g: Survey and challenges,'' \emph{IEEE Communications Magazine},
  vol.~55, no.~5, pp. 94--100, 2017.

\bibitem{SAA}
A.~J. Kleywegt, A.~Shapiro, and T.~Homem-de Mello, ``The sample average
  approximation method for stochastic discrete optimization,'' \emph{SIAM
  Journal on Optimization}, vol.~12, no.~2, pp. 479--502, 2002.

\bibitem{LawofLargeNumbers}
C.~Robert and G.~Casella, \emph{Monte Carlo statistical methods}.\hskip 1em
  plus 0.5em minus 0.4em\relax Springer Science \& Business Media, 2013.

\bibitem{searching-book}
E.~K. Burke, G.~Kendall \emph{et~al.}, \emph{Search methodologies}.\hskip 1em
  plus 0.5em minus 0.4em\relax Springer, 2005.

\bibitem{GA-complexity}
B.~L. Miller, D.~E. Goldberg \emph{et~al.}, ``Genetic algorithms, tournament
  selection, and the effects of noise,'' \emph{Complex systems}, vol.~9, no.~3,
  pp. 193--212, 1995.

\bibitem{composition3}
S.~{Deng}, H.~{Wu}, W.~{Tan}, Z.~{Xiang}, and Z.~{Wu}, ``Mobile service
  selection for composition: An energy consumption perspective,'' \emph{IEEE
  Transactions on Automation Science and Engineering}, vol.~14, no.~3, pp.
  1478--1490, July 2017.

\bibitem{review-add2}
H.~{Yuan}, J.~{Bi}, and M.~{Zhou}, ``Temporal task scheduling of multiple
  delay-constrained applications in green hybrid cloud,'' \emph{IEEE
  Transactions on Services Computing}, pp. 1--1, 2018.

\bibitem{review-add4}
Q.~{Wu}, M.~{Zhou}, Q.~{Zhu}, and Y.~{Xia}, ``Vcg auction-based dynamic pricing
  for multigranularity service composition,'' \emph{IEEE Transactions on
  Automation Science and Engineering}, vol.~15, no.~2, pp. 796--805, 2018.

\bibitem{provision1}
H.~Wu, S.~Deng, W.~Li, J.~Yin, Q.~Yang, Z.~Wu, and A.~Y. Zomaya,
  ``Revenue-driven service provisioning for resource sharing in mobile cloud
  computing,'' in \emph{Service-Oriented Computing}, 2017, pp. 625--640.

\bibitem{discovery1}
W.~{Chen}, I.~{Paik}, and P.~C.~K. {Hung}, ``Constructing a global social
  service network for better quality of web service discovery,'' \emph{IEEE
  Transactions on Services Computing}, vol.~8, no.~2, pp. 284--298, March 2015.

\bibitem{review-add3}
Q.~{Fan} and N.~{Ansari}, ``On cost aware cloudlet placement for mobile edge
  computing,'' \emph{IEEE/CAA Journal of Automatica Sinica}, vol.~6, no.~4, pp.
  926--937, 2019.

\bibitem{redundancy--1}
H.~{Deng}, T.~{Zhao}, and I.~{Hou}, ``Online routing and scheduling with
  capacity redundancy for timely delivery guarantees in multihop networks,''
  \emph{IEEE/ACM Transactions on Networking}, vol.~27, no.~3, pp. 1258--1271,
  June 2019.

\bibitem{redundancy--2}
A.~{abdi}, A.~{Girault}, and H.~R. {Zarandi}, ``Erpot: A quad-criteria
  scheduling heuristic to optimize execution time, reliability, power
  consumption and temperature in multicores,'' \emph{IEEE Transactions on
  Parallel and Distributed Systems}, vol.~30, no.~10, pp. 2193--2210, Oct 2019.

\bibitem{redundancy--3}
H.~{Xu}, G.~{De Veciana}, W.~C. {Lau}, and K.~{Zhou}, ``Online job scheduling
  with redundancy and opportunistic checkpointing: A speedup-function-based
  analysis,'' \emph{IEEE Transactions on Parallel and Distributed Systems},
  vol.~30, no.~4, pp. 897--909, April 2019.

\bibitem{similar}
Y.~Chen, S.~Deng, H.~Ma, and J.~Yin, ``Deploying data-intensive applications
  with multiple services components on edge,'' \emph{Mobile Networks and
  Applications}, Apr 2019.

\end{thebibliography}

\begin{IEEEbiography}
  [{\includegraphics[width=1in,height=1.25in,clip,keepaspectratio]{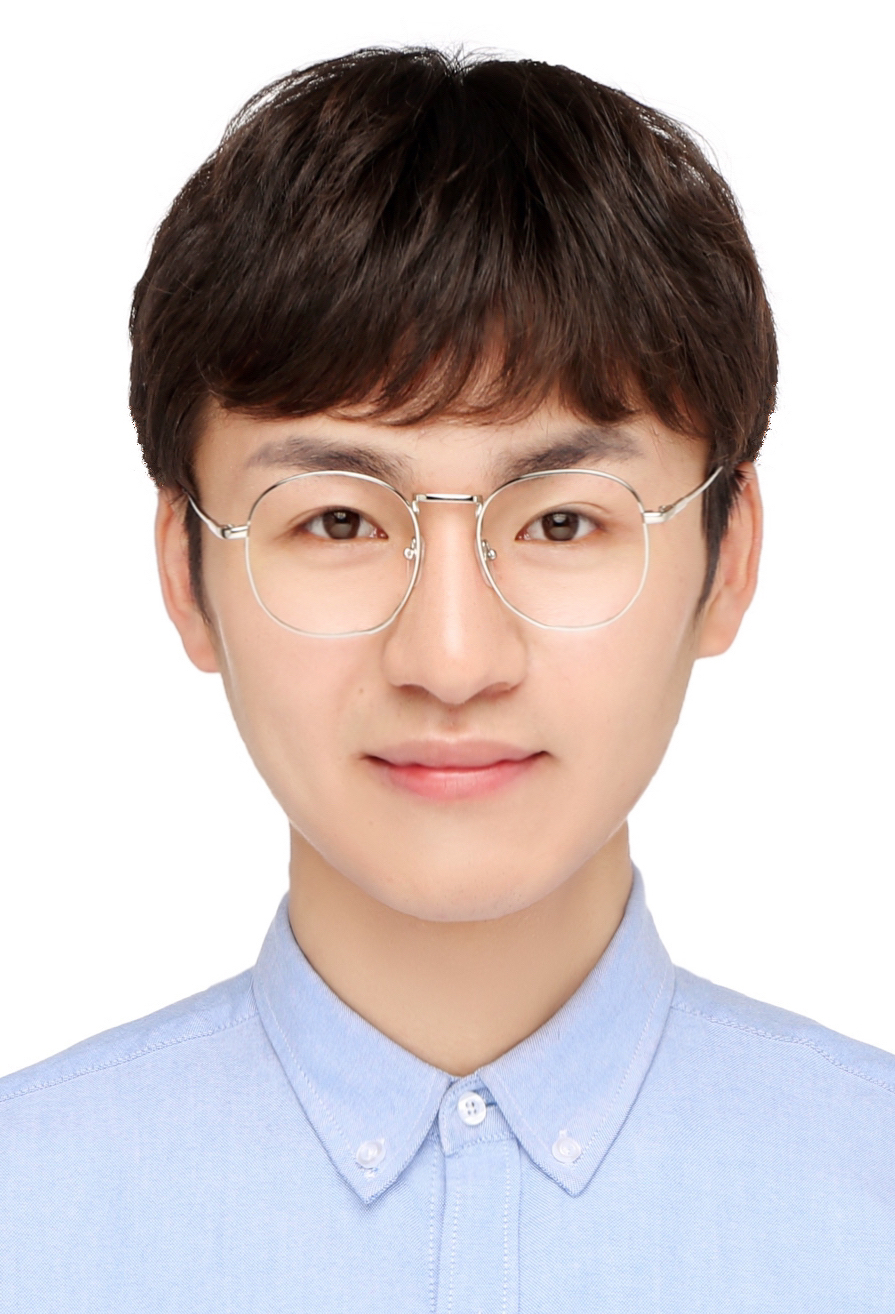}}]{Hailiang Zhao} 
  received the B.S. degree in 2019 from the school of computer science and technology, 
  Wuhan University of Technology, Wuhan, China. He is currently pursuing the Ph.D. degree with the 
  College of Computer Science and Technology, Zhejiang University, Hangzhou, China. He has been a 
  recipient of the Best Student Paper Award of IEEE ICWS 2019. His research interests include edge 
  computing, service computing and machine learning.
\end{IEEEbiography}

\begin{IEEEbiography}
  [{\includegraphics[width=1in,height=1.25in,clip,keepaspectratio]{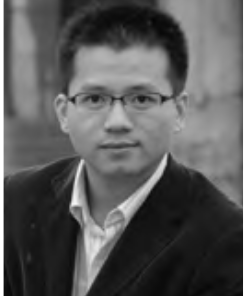}}]{Shuiguang Deng} 
  is currently a full professor at the College of Computer Science and Technology 
  in Zhejiang University, China, where he received a BS and PhD degree both in Computer Science in 
  2002 and 2007, respectively. He previously worked at the Massachusetts Institute of Technology in 
  2014 and Stanford University in 2015 as a visiting scholar. His research interests include Edge 
  Computing, Service Computing, Mobile Computing, and Business Process Management. He serves as the 
  associate editor for the journal IEEE Access and IET Cyber-Physical Systems: Theory \& Applications. 
  Up to now, he has published more than 100 papers in journals and refereed conferences.  In 2018, he 
  was granted the Rising Star Award by IEEE TCSVC. He is a fellow of IET and a senior member of IEEE.
\end{IEEEbiography}

\begin{IEEEbiography}
  [{\includegraphics[width=1in,height=1.25in,clip,keepaspectratio]{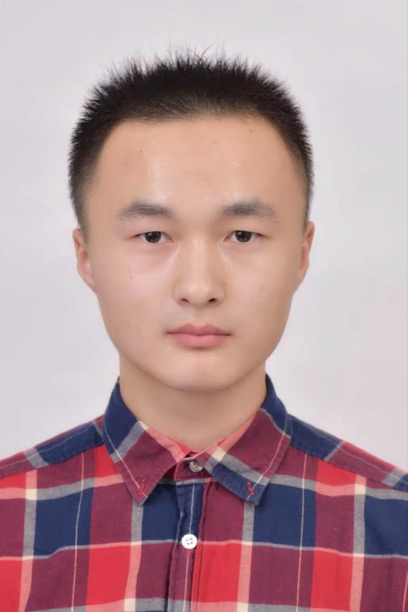}}]{Zijie Liu} 
  received the B.S. degree in 2018 from the school of computer science and technology, 
  Huazhong University of Science an Technology, Wuhan, China. He is now pursuing the master degree 
  with the College of Computer Science and Technology, Zhejiang University, Hangzhou, China. His 
  research interests include edge computing and software engineering.
\end{IEEEbiography}

\begin{IEEEbiography}
  [{\includegraphics[width=1in,height=1.25in,clip,keepaspectratio]{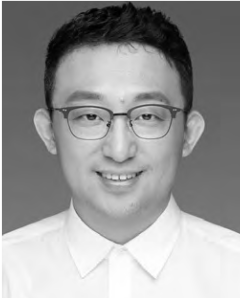}}]{Jianwei Yin} 
  received the Ph.D. degree in computer science from Zhejiang University (ZJU) in 2001. 
  He was a Visiting Scholar with the Georgia Institute of Technology. He is currently a Full Professor 
  with the College of Computer Science, ZJU. Up to now, he has published more than 100 papers in top 
  international journals and conferences. His current research interests include service computing 
  and business process management. He is an Associate Editor of the IEEE Transactions on Services 
  Computing.
\end{IEEEbiography}

\begin{IEEEbiography}[{\includegraphics[width=1in,height=1.25in,clip,keepaspectratio]{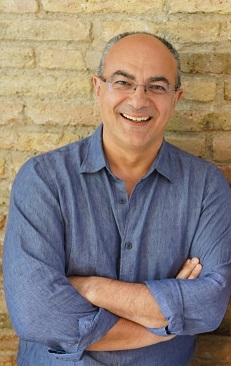}}]{Schahram Dustdar}
  is a Full Professor of Computer Science (Informatics) with a focus on Internet Technologies heading the Distributed 
  Systems Group at the TU Wien. He is Chairman of the Informatics Section of the Academia Europaea (since December 9, 2016). 
  He is elevated to IEEE Fellow (since January 2016). From 2004-2010 he was Honorary Professor of Information Systems 
  at the Department of Computing Science at the University of Groningen (RuG), The Netherlands.

  From December 2016 until January 2017 he was a Visiting Professor at the University of Sevilla, Spain and from January 
  until June 2017 he was a Visiting Professor at UC Berkeley, USA. He is a member of the IEEE Conference Activities Committee 
  (CAC) (since 2016), of the Section Committee of Informatics of the Academia Europaea (since 2015), a member of the Academia 
  Europaea: The Academy of Europe, Informatics Section (since 2013). He is recipient of the ACM Distinguished Scientist award 
  (2009) and the IBM Faculty Award (2012). He is an Associate Editor of IEEE Transactions on Services Computing, ACM Transactions 
  on the Web, and ACM Transactions on Internet Technology and on the editorial board of IEEE Internet Computing. He is the 
  Editor-in-Chief of Computing (an SCI-ranked journal of Springer).
\end{IEEEbiography}

\end{document}